%% file: master.tex
\pdfoutput=1

\documentclass[10pt,journal]{IEEEtran}
\input{acronyms.tex}

\usepackage{cite}
\ifCLASSINFOpdf
  \usepackage[pdftex]{graphicx}
  \graphicspath{{./figures/}}
  \DeclareGraphicsExtensions{.pdf}
\else
\fi
\usepackage{color}

\usepackage[section]{placeins}

\usepackage{amsfonts}
\usepackage{amsmath}
\input{commands.tex}

\usepackage{algpseudocode}%algorithmicx
\usepackage{array}
\ifCLASSOPTIONcompsoc
  usepackage[caption=false,font=normalsize,labelfont=sf,textfont=sf]{subfig}
\else
  \usepackage[caption=false,font=footnotesize]{subfig}
\fi

\newcommand{\tcb}{\textcolor{blue}}

\begin{document}
\title{Hardware-Based Linear Program Decoding with the Alternating Direction Method of Multipliers}

\author{Mitchell Wasson, Mario
  Milicevic,~\IEEEmembership{Student~Member,~IEEE}, Stark
  C. Draper,~\IEEEmembership{Senior~Member,~IEEE}, \\ and Glenn
  Gulak,~\IEEEmembership{Senior~Member,~IEEE}% <-this % stops a space
  \thanks{This material was presented in part at the 2015 Asilomar
    Conf.~on Signals, Systems, and Computers, Pacific Grove, CA, Nov. 2015.}
\thanks{M.~Wasson was with, and M.~Milicevic, S.~Draper, and
  G.~Gulak are with, the Dept.~of Electrical and Comp. Eng.,
  University of Toronto, ON M5S 3G4, Canada (e-mail:
  m.wasson@mail.utoronto.ca, mario.milicevic@utoronto.ca,
  stark.draper@utoronto.ca, gulak@eecg.toronto.edu)}
% 
%\thanks{M.~Wasson was with the Dept.~of Electrical and Computer
%  Eng., University of Toronto, ON M5S 3G4, Canada (email: m.wasson@mail.utoronto.ca).}
% 
%\thanks{M.~Milicevic is with the Dept.~of Electrical and Computer
%Eng., University of Toronto, ON M5S 3G4, Canada (e-mail: mario.milicevic@utoronto.ca)}
%
%\thanks{S.~C.~Draper is with the Dept.~of Electrical and Computer
%Eng., University of Toronto, ON M5S 3G4, Canada (e-mail:
%stark.draper@utoronto.ca)}
% 
%\thanks{G.~Gulak is with the Dept.~of Electrical and Computer
%  Eng., University of Toronto, ON M5S 3G4, Canada (e-mail:
%  gulak@eecg.toronto.edu)}
% 
\thanks{This work was supported by the National Science Foundation
(NSF) under Grant CCF-1217058, by the Natural Science and
Engineering Research Council (NSERC) of Canada, including through a
Discovery Research Grant, and by the Canadian Microelectronics
Corporation (CMC).}  
%\thanks{Thanks to someone?}% <-this % stops a space
%\thanks{manuscript received stuff...}%
}


%\markboth{submitted to IEEE TRANSACTIONS ON SIGNAL PROCESSING, November 2016} %, VOL. XX, NO. X, MONTH 201X}
{}



%counter used for algorithms
%need this since algorithms have to occur in the figure environment
\newcounter{AlgorithmCounter}
\setcounter{AlgorithmCounter}{0}
\newcounter{FigureCounter}
\setcounter{FigureCounter}{0}
\setcounter{figure}{\value{FigureCounter}}

\maketitle

\begin{abstract}
We present a hardware-based implementation of \gls{lp} decoding for
binary linear codes.  \gls{lp} decoding frames error-correction as an
optimization problem. In contrast, variants of \gls{bp} decoding frame
error-correction as a problem of graphical inference.  \gls{lp}
decoding has several advantages over \gls{bp}-based methods, including
convergence guarantees and better error-rate performance in
high-reliability channels.  The latter makes \gls{lp} decoding
attractive for optical transport and storage applications.  However,
\gls{lp} decoding, when implemented with general solvers, does not
scale to large blocklengths and is not suitable for a parallelized
implementation in hardware.  It has been recently shown that the
\gls{admm} can be applied to decompose the \gls{lp} decoding problem.
The result is a message-passing algorithm with a structure very
similar to \gls{bp}.  We present new intuition for this decoding
algorithm as well as for its major computational primitive: projection
onto the parity polytope.  Furthermore, we present results for a
fixed-point Verilog implementation of \admmlp decoding.  This
implementation targets a \gls{fpga} platform to evaluate error-rate
performance and estimate resource usage.  We show that \gls{fer}
performance well within 0.5dB of double-precision implementations is
possible with 10-bit messages.  Finally, we outline a number of
research opportunities that should be explored en-route to the
realization of an \gls{asic} implementation capable of gigabit per
second throughput.
\end{abstract}

%\begin{IEEEkeywords}
%\end{IEEEkeywords}

% \fi
\IEEEpeerreviewmaketitle

%reset the acronym display
\glsresetall

%here is where the actual article happens
\section{Introduction}
\input{introduction.tex}

\section{Background}
\input{background.tex}

\section{Algorithms}
\label{algSec}
\input{admmDecomposition.tex}
\input{parityPolytope.tex}

\input{simplex.tex}

\section{Hardware Architecture}

\input{decoderArch.tex}
\input{variableNode.tex}
\input{checkNode.tex}
\input{ppArch.tex}
\input{simpArch.tex}

\section{Results}
\input{results.tex}

\input{errorResults.tex}

\input{resourceResults.tex}

\input{implementationComp.tex}

\section{Discussion and Conclusions}
\input{conclusion.tex}
\ifCLASSOPTIONcaptionsoff
  \newpage
\fi



\bibliographystyle{IEEEtran}
\bibliography{IEEEabrv,master}
\end{document}

%% file: acronyms.tex
\usepackage[acronym,nonumberlist]{glossaries}
\newacronym{admm}{ADMM}{Alternating Direction Method of Multipliers}
\newacronym{alm}{ALM}{Adaptive Logic Modules}
\newacronym{ber}{BER}{Bit Error Rate}
\newacronym{asic}{ASIC}{Application Specific Integrated Circuit}
\newacronym{lp}{LP}{Linear Program}
\newacronym{ldpc}{LDPC}{Low-Density Parity-Check}
\newacronym{ml}{ML}{Maximum Likelihood}
\newacronym{qc}{QC}{Quasi-Cyclic}
\newacronym{fer}{FER}{Frame Error Rate}
\newacronym{fpga}{FPGA}{Field-Programmable Gate Array}
\newacronym{bp}{BP}{Belief Propagation}
\newacronym{pc}{PC}{Personal Computer}
\newacronym{awgn}{AWGN}{Additive White Gaussian Noise}
\newacronym{llr}{LLR}{Log-Likelihood Ratio}
\newacronym{kkt}{KKT}{Karush-Kuhn-Tucker}
\newacronym{snr}{SNR}{Signal-to-Noise Ratio}
\newacronym{vn}{VN}{Variable Node}
\newacronym{cn}{CN}{Check Node}
\newacronym{ctv}{CN-to-VN}{Check-to-Variable}
\newacronym{vtc}{VN-to-CN}{Variable-to-Check}
\newacronym{dsp}{DSP}{Digital Signal Processing}
\newacronym{ram}{RAM}{Random Access Memory}

\makeglossaries

%% file: commands.tex
\newcommand{\card}[1]{\left\vert{#1}\right\vert}
\newcommand{\norm}[2]{{\left\lVert{#1}\right\rVert}_{#2}}
\newcommand{\admmlp}[0]{\gls{admm}-\gls{lp} }
\newcommand{\qcldpc}[0]{\gls{qc}-\gls{ldpc} }
\newcommand{\ceil}[1]{\left\lceil{#1}\right\rceil}
\newcommand{\bigO}[1]{\mathcal{O}\left({#1}\right)}

\newcommand{\AlgCaption}[1]{\renewcommand{\figurename}{Alg.}\setcounter{figure}{\value{AlgorithmCounter}}\caption{#1}\stepcounter{AlgorithmCounter}\renewcommand{\figurename}{Fig.}\setcounter{figure}{\value{FigureCounter}}}

\newcommand{\FigCaption}[1]{\caption{#1}\stepcounter{FigureCounter}}

\usepackage{multirow}
\usepackage{booktabs}
\usepackage{makecell}
\newcolumntype{L}[1]{>{\raggedright\let\newline\\\arraybackslash\hspace{0pt}}m{#1}}
\newcolumntype{C}[1]{>{\centering\let\newline\\\arraybackslash\hspace{0pt}}m{#1}}
\newcolumntype{R}[1]{>{\raggedleft\let\newline\\\arraybackslash\hspace{0pt}}m{#1}}

%% file: introduction.tex
%\section{Introduction}

\IEEEPARstart{T}{he} field of error-correction coding was
revolutionized in the mid-1990s by the widespread adoption (and
academic study) of graph-based codes and associated \gls{bp} message-passing decoding
algorithms~\cite{berrou_1993_turbo,mackay_1995_ldpc,kschischang_2001_sum_product}.
A key aspect of the success of these codes was their compatibility
with hardware.  \gls{bp}-based decoders are naturally distributed
algorithms and variants such as Min-Sum are (relatively) easily mapped
to hardware.  Graph-based codes, particularly Turbo codes and
\gls{ldpc} codes, have been adopted in many real world systems.
However, there are issues present with \gls{bp}-based decoding
algorithms.  The first is their reliance on the tree assumption for
the code-defining graph.  In practice tree codes are not used due to
their poor distance properties.  This results in the use of \gls{ldpc}
codes without performance or convergence guarantees due to graph cycles.
Additionally, it is observed in practice that \gls{bp}-based decoding
algorithms often suffer from performance deficiencies, termed ``error
floors'', in high-reliability channels~\cite{richardson_2003_error}.

In the early 2000s, Feldman and his collaborators realized that the
\gls{ml} decoding problem for binary linear codes can be rephrased as
an integer program~\cite{feldman_2005_journal}.  One obtains a 
\gls{lp} by relaxing the integer constraints.  Feldman's
work applies to any binary linear code, but he concentrated on
\gls{ldpc} codes due to their prevalence and smaller constraint sets.
These results generated much interest among coding theorists.
\gls{lp}s are an extremely well-studied and understood class of
optimization problems, especially when contrasted with \gls{bp}.  For
instance, \gls{lp} decoding has an \gls{ml} certificate
property~\cite{feldman_2005_journal}, such that if \gls{lp} decoding fails, it
fails in a detectable way (to a non-integer vertex). The relaxation
can then be tightened and the \gls{lp}
re-run~\cite{seigel_2008_adaptive_lp}.  If a high-quality expander or
high-girth code is used, \gls{lp} decoding is guaranteed to correct a
constant number of bit
flips~\cite{feldman_2005_constant,arora_2009_constant}.  Broadly, it
was hoped that by studying \gls{lp} decoding, more would be understood
about \gls{bp} decoding.

On the practical side, there was less excitement.  There initially
seemed to be no real-world need for such a decoder.  Furthermore,
traditional \gls{lp} solvers did not scale easily to the blocklengths
of modern error-correcting codes.  Nevertheless, a number of groups
did study how to build an application-specific low-complexity \gls{lp}
decoder~\cite{vontobel_2006_low_complex_lp,burshtein_2009_iterative_lp,seigel_2008_adaptive_lp,draper_2013_admm_lp}.
In particular, Barman et al.~\cite{draper_2013_admm_lp} built an
application-specific \gls{lp} decoder that was computationally
competitive with \gls{bp} and that had a message-passing structure
with a standard schedule \cite{draper_2013_admm_lp}.  They solved
the \gls{lp} decoding problem using the \gls{admm}, a decomposition technique used in
large-scale optimization~\cite{boyd_2011_admm}.  Able to study \gls{lp} decoding
performance at long blocklengths, it was observed empirically, and
later confirmed theoretically, that \gls{lp} decoding far outperforms
\gls{bp} in the high \gls{snr}
regime~\cite{draper_2013_admm_lp,liu_2014_instanton,liu_2015_jumpLinear}.
In this regime, \gls{lp} decoders do not suffer from the same error
floor effects as \gls{bp}.  Using the \gls{admm} solver, Liu and
Draper were able to augment the objective of \gls{lp} decoding with a
penalty term to improve error-rates further in low-reliability
channels~\cite{draper_2014_penalized_lp}.  Additionally, \gls{lp}
decoding can be used as a subroutine in a multi-stage decoder that
quickly approaches \gls{ml} performance~\cite{wang_2009_multistage}.
Thus, for applications in which reliability demands are extreme,
\gls{lp} decoding is an attractive alternative (or complement) to
\gls{bp}.  Further generalization of ADMM-LP to non-binary and 
multipermutation codes are developed
in~\cite{liu_2016_nonBinADMM,liu_2016_multiperm}.

In parallel to these theoretical and algorithmic developments, there
has been growing interest in moving \admmlp toward a hardware
implementation.  Several groups have made progress in creating
efficient methods for solving the key computational primitive
of \admmlp decoding, Euclidean projection onto the ``parity
polytope''~\cite{siegel_2013_projection_lp,kleijn_2013_pp_projection,wasson_2015_pp}.
In particular, Wasson and Draper investigated mapping this operation
to hardware~\cite{wasson_2015_pp}.  Several implementation papers have
also considered ADMM-LP decoding in other contexts.  Debbabi et
al.~investigated how to schedule messages more efficiently and
developed a multicore implementation~\cite{debbabi2016,debbabi2015}.
Jiao et al.~modified penalization to improve error-rate performance of
irregular LDPC codes~\cite{Jiao2015}.  Finally, Wei et al.~implemented
ADMM-LP avoiding projections when possible~\cite{wei2016}.

%However, one major hurdle remains that will determine whether or
%not \gls{admm} is truly a viable competitor to \gls{bp} in high-reliability
%applications. That hurdle is to show that \admmlp decoding
%algorithms can be mapped to hardware without unacceptable performance loss.

While useful investigations, these studies do not demonstrate whether
or not ADMM-LP decoding is viable in hardware.  In
this paper, we present a \gls{fpga}-based implementation that shows
that the \admmlp decoding algorithm can be mapped to hardware without
suffering an unacceptable performance loss.  First, we review and
expand upon the execution and intuition of the \admmlp decoding
algorithm.  Next, we review the developments made
in~\cite{wasson_2015_pp} to implement Euclidean projection onto the
parity polytope in hardware.  We then describe how to assemble the
pieces to form a complete LP decoder.  We present results for the
[155, 64] \gls{qc} \gls{ldpc} code introduced by Tanner et
al.~\cite{tanner_2001_refcode}, the [672, 546] \qcldpc code from the
IEEE 802.11ad (WiGig) standard~\cite{IEEE-80211ad_rev}, and an ensemble of
(3,6)-regular [1002, 503] \qcldpc codes.  We test code performance
using an \gls{fpga}-based simulation environment.  While our
initial implementation requires more hardware resources than Min-Sum decoders, we find that it is possible to preserve the superior error-rate performance of ADMM-LP in fixed point.

%% file: background.tex
%\section{Background}

In this paper, we consider the decoding of binary linear codes. A
binary linear code $\mathcal{C}$ of blocklength $n$ is a
$k$-dimensional subspace of $\mathbb{F}_2^n$.  Such a code can be
defined as the null space of the $m\times n$ ``parity-check'' matrix
$H$, i.e., $\mathcal{C} = \left\{ x \in \{0,1\}^n : Hx =
0 \pmod{2} \right\}$.  In general $m \geq n - k$ with equality when
$H$ has full rank.  The rate of $\mathcal{C}$ is defined to be $R=k/n$
which specifies the number of information bits transmitted per
codeword symbol. Each row of the parity-check matrix corresponds to a
check, which specifies a subset of codeword symbols that must add to 0
modulo 2. These checks are indexed by the set $\mathcal{J}
= \{1, \dots, m\}$. Each column of the parity-check matrix corresponds
to a codeword symbol or variable, indexed by $\mathcal{I}
= \{1, \dots, n\}$. The neighborhood of check $j$, denoted
$\mathcal{N}_c\left(j\right)$, is the set of variables that check $j$
constrains to add to 0. That is, $\mathcal{N}_c\left(j\right) = \{ i \in \mathcal{I} :
H_{j,i} = 1 \}$. Similarly, the neighborhood of variable $i$,
$\mathcal{N}_v\left(i\right)$, is the set of checks in which variable
$i$ participates, $\mathcal{N}_v\left(i\right) = \left\{
j\in\mathcal{J} : H_{j,i} = 1
\right\}$.  %A binary vector $x \in \mathcal{C}$ if
%and only if it satisfies each check.

Given a stochastic channel model $\mathcal{P}(y|x)$ where
$y\in\mathcal{Y}^n$ is the channel output, \gls{ml} decoding amounts to 
maximizing the model over the set of codewords.
That is, we decode to $\arg
\max_{x\in \mathcal{C}}\mathcal{P}(y|x)$. % with tie-breaking in case
%the maximizer is not unique.  
It was shown in~\cite{feldman_2003_phd} that, when considering a
binary linear code transmitted over a symmetric memoryless channel,
the \gls{ml} decoding objective is linear in the length-$n$ vector
$\gamma$ of Log-likelihood Ratios (LLRs) $\gamma_i
= \log(\mathcal{P}(y_i|0) / \mathcal{P}(y_i | 1) )$ .  \gls{ml}
decoding problem thus is 
\begin{equation}
\arg \max_{x \in \mathcal{C}} \mathcal{P}(y|x) = \arg \min_{x \in \mathcal{C}} \sum_{i=1}^n \gamma_i x_i = \arg \min_{x \in \mathcal{C}} \gamma^T x. \label{MLdef}
\end{equation}
We note that $\gamma$ can be multiplied by any positive scalar without
changing the problem.

%
%where $\gamma_i = \log \left(
%\frac{\mathcal{P}(y_i|0)}{\mathcal{P}(y_i | 1)}\right)$.  
%The equivalence is quick to show:
%\begin{align*}
%\log \mathcal{P}(y|x) &= \log \prod_{i=1}^n \mathcal{P} (y_i|x_i) =
%\sum_{i=1}^n \log \mathcal{P} (y_i|x_i)\\ & = \sum_{i=1}^n
%\left[\log\left(\frac{\mathcal{P}(y_i|1)}{\mathcal{P}(y_i|0)}\right)
%  x_i + \mathcal{P}(y_i|0) \right].
%\end{align*}
%And so,
%\begin{equation}
%\arg \max_{x \in \mathcal{C}} \mathcal{P}(y|x) = \arg \min_{x \in \mathcal{C}} \sum_{i=1}^n \gamma_i x_i = \arg \min_{x \in \mathcal{C}} \gamma^T x, \label{MLdef}
%\end{equation}
%where $\gamma$ is the length-$n$ vector of the $\gamma_i$.  
%Note that $\gamma$ can be multiplied by any positive scalar without
%changing the problem.  We comment that \gls{ml} decoding minimizes the
%frame error rate (\gls{fer}) for a uniform prior over codewords.  This
%contrasts with \gls{bp} which attempts to minimize the symbol error
%rate (bit error rate for binary codes) by approximating the
%symbol-wise marginal distributions of $\mathcal{P}(y|x)$.

Having framed \gls{ml} as an optimization problem with a linear
objective, we are ready to develop the \gls{lp} relaxation first
proposed in~\cite{feldman_2003_phd}.  First, denote by
$x_{\mathcal{S}}$, $\mathcal{S} \subseteq \mathcal{I}$, the
length-$|\mathcal{S}|$ vector formed with the components of $x$
indexed by $\mathcal{S}$. With this notation, we can restate the
parity-check condition for a valid codeword as $\mathcal{C} = \left\{
x \in \{0,1\}^n : 1^\top x_{\mathcal{N}_c\left(j\right)} = 0 \pmod{2}
\text{ for all } j\in\mathcal{J} \right\}$.  Each of the $m$
constraints in this set can be visualized as requiring that the set of
codeword variables connected to any particular check must be an
even-weight vertex of the unit hyper-cube.

%\footnote{We note that in
%this visualization we have used a seemingly natural embedding of
%$\mathbb{F}_2$ into $\mathbb{R}$: we map $0 \in \mathbb{F}_2$ to
%$0 \in \mathbb{R}$ and $1 \in \mathbb{F}_2$ to $1 \in \mathbb{R}$.  We
%needed to embed the finite field over which the code is defined in
%order to apply optimization techniques. If, however, one considers
%embeddings of non-binary codes (defined over fields other than
%$\mathbb{F}_2$), then one quickly realizes that the question of how to
%embed a finite field in the reals is generally subtle and involved.
%See, e.g., \cite{flanagan_09_lp,liu_2016_nonBinADMM} for details.}

\begin{figure}%[h]
	\centering \includegraphics[width=0.2\textwidth]{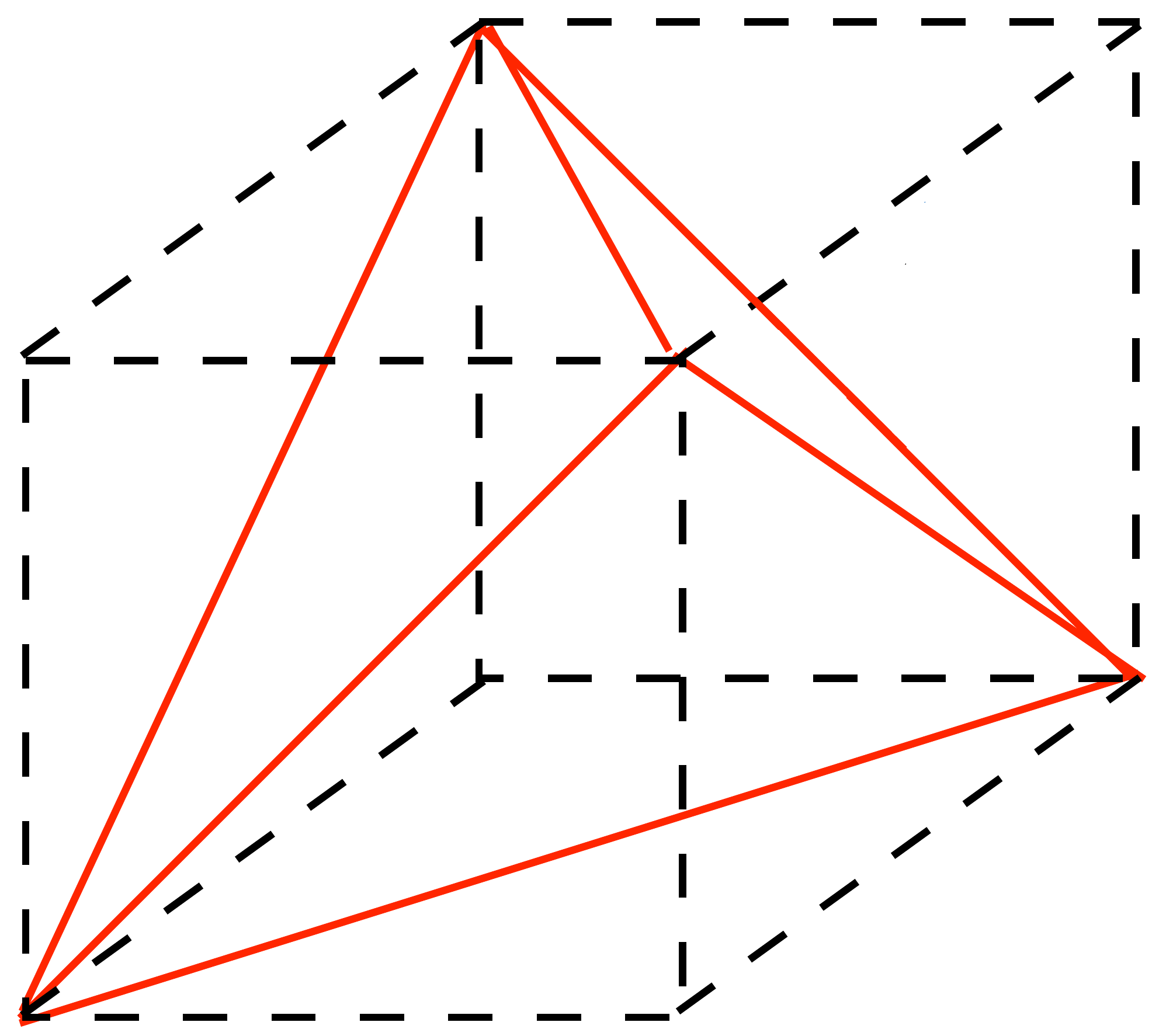}
        \FigCaption{Visualization of $\mathbb{PP}_3$ within the
          three-dimensional unit cube.}
	\label{pp3Picture}
\end{figure}

The LP decoding problem results from relaxing these constraints
\cite{feldman_2003_phd, feldman_2005_journal}. Instead of requiring
the vector of variables connected to each check to be an even-weight
vertex of the unit hyper-cube, LP decoding rather requires this set of
variables to lie in the convex hull of the vertices.  Visualized in
Fig.~\ref{pp3Picture}, the convex hull of the even-weight vertices of
the unit hyper-cube is termed the ``parity polytope'', denoted
$\mathbb{PP}_d$ in $d$-dimension:
\begin{equation}
\label{PPdef}
\mathbb{PP}_d := \text{conv}\left(\left\{ e \in \left\{ 0,1 \right\}^d
: 1^\top e = 0 \pmod{2} \right\}\right).
\end{equation}

The polytope $\mathbb{PP}_d$ can be explicitly defined by a number of
half-space inequalities~\cite{feldman_2003_phd}.  Every odd-weight
vertex is surrounded by even-weight vertices.  Each half-space
inequality is defined by the hyperplane that contains all these
even-weight vertices and ``cuts'' off the half of the space in which
the odd-weight vertex sits.  Half of the $2^d$ vertices are of odd
weight, so we can describe $\mathbb{PP}_d$ with $2^{d-1}$ half-space
constraints.  Each such inequality corresponds to one of the
constraints in the first line of the following description of
$\mathbb{PP}_d$ where we use the notation $[d]=\left\{
1,\dots,d \right\}$.  A vector $v \in
\mathbb{PP}_d$ if
\begin{equation}
	\label{ppCutDef}
	{\begin{array}{ll}	
		\sum\limits_{i \in S} v_i - \sum\limits_{i \in[d] \backslash \mathcal{S}} v_i \leq \card{\mathcal{S}} - 1 & \quad \mathcal{S} \subseteq [d], \text{ } \card{\mathcal{S}} \text{ odd}\vspace{0.5ex}\\
0 \leq v_i \leq 1 & \quad i\in[d] 
	\end{array}}.
\end{equation}
The box constraints $0 \leq v_i \leq 1$ are not always
redundant, e.g., when $d=2$.  

%The number of linear constraints in~(\ref{ppCutDef}) is exponential in
%$d$.  More efficient descriptions with a quadratic number of
%constraints do exist~\cite{yannakakis_1991_pp}.  However, the
%exponential description complexity will not prove limiting because, as
%we will see, our main need will be to develop algorithms that project
%onto $\mathbb{PP}_d$ and, if one is clever, one can accomplish
%projection without considering all $2^{d-1}$ constraints.

In summary, \gls{lp} decoding requires us to solve 
\begin{equation}
\label{LPDecoding}
{\begin{array}{rll}
		\arg \min\limits_{x} & \gamma^\top x\\
		\text{subject to}& x_{\mathcal{N}_c\left(j\right)} \in \mathbb{PP}_{\left\vert{\mathcal{N}_c(j)}\right\vert}  & j\in\mathcal{J} \vspace{1ex}\\
		& x \in \left[0,1\right]^n 
	\end{array}}
\end{equation}
%where $x$ is also constrained to be in the unit cube, though often
%this constraint is redundant.  
%\tcr{As each parity-check constrains a
%subset of variables to be in a convex set, the unit cube is a convex
%set, and the intersection of a collection of convex sets is a convex
%set, the feasible set is convex.  Thus, the \gls{lp} has no local
%minima.}  
Note that \gls{lp} decoding is not guaranteed to yield the \gls{ml}
solution.  Due to the relaxation, the feasible space has fractional
vertices.  The failure model of \gls{lp} decoding is when one of these
``pseudocodewords'' is the minimal cost vertex.

%We now make some observations about \gls{lp} decoding.
%In~(\ref{LPDecoding}) we optimize the same linear objective
%as \gls{ml}~(\ref{MLdef}) over a feasible set containing all
%codewords.  Therefore, if the solution to the \gls{lp} is a codeword,
%it must be the \gls{ml} decoding.  Furthermore, the only all-binary
%vectors contained in the feasible set are codewords and thus the
%result of
%\gls{lp} decoding is the \gls{ml} decoding result if it is a binary
%vector.  This is the ``ML certificate''
%property~\cite{feldman_2003_phd}.  However, \gls{lp} decoding is not
%guaranteed to give the \gls{ml} decoding.  Due to the relaxation, the
%feasible space has fractional vertices.  The failure model of \gls{lp}
%decoding is when one of these ``pseudocodewords'' is the minimal cost
%vertex.

One might think of rounding the fractional components when
the \gls{lp} solver outputs a pseudocodeword. However, this does not
solve the pseudocodewords problem.  An alternative approach proposed
by Liu and Draper~\cite{draper_2014_penalized_lp} is to augment the
objective of (\ref{LPDecoding}) with a penalty function.  This
approach, referred to as ``penalized \gls{lp}'' decoding, discourages
pseudocodewords by penalizing the closeness of variables to $1/2$.
Many penalty functions were tested in~\cite{draper_2014_penalized_lp},
but we only implement the so-called $\ell_1$-penalty function, due to
its good error-rate performance and algorithmic simplicity.  The
$\ell_1$-penalized
\gls{lp} decoding problem is given by
\begin{equation}
\label{penalizedLPDecoding}
{\begin{array}{r l l}
	\min & \gamma^\top x -\alpha\norm{x-\frac{1}{2}}{1}\\
	\text{subject to} & x_{\mathcal{N}_c(j)} \in \mathbb{PP}_{\card{\mathcal{N}_c(j)}} & j\in\mathcal{J} \vspace{1ex}\\
		& x \in \left[0,1\right]^n 
	\end{array}}
\end{equation}
where $\alpha\geq 0$ is termed the penalty parameter.  The penalty
parameter tunes how severely non-binary variables should be penalized.
Setting $\alpha=0$ reduces~(\ref{penalizedLPDecoding})
to~(\ref{LPDecoding}).  While moderate values of $\alpha$ improve
performance in the waterfall regime, an excessively large $\alpha$ can
adversely affect performance~\cite{draper_2014_penalized_lp}.

%The addition of the penalty function makes the objective of
%(\ref{penalizedLPDecoding}) non-convex.  But, as we will see, when
%using \gls{admm} the non-convexity does not pose an algorithmic issue.
%It does, however, remove the certificate of correctness.  Empirically,
%penalizing the objective can dramatically reduce the occurrence of error
%events caused by pseudocodewords~\cite{draper_2014_penalized_lp}.

Up until this point, we have been discussing prior formulations of \gls{lp}
decoding.  We now discuss a simple transformation of \gls{lp} decoding
that proves useful when designing a fixed-point implementation.  The
original \gls{lp} decoding formulation operates inside the unit
hypercube centered around $1/2$.  In our hardware implementation,
signed integers are used to implement fixed-point arithmetic.
Therefore, to eliminate possible asymmetries, we prefer \gls{lp}
decoding to operate inside the unit hypercube symmetrically centered
around $0$. To accomplish this, the simple
variable substitution $x_{\text{new}}=x_{\text{old}}-1/2$ can be
applied to (\ref{penalizedLPDecoding}).  The result
\begin{equation}
\label{centeredLPDecoding}
{\begin{array}{r l l}
	\min & \gamma^\top x -\alpha\norm{x}{1}\\
	\text{subject to} & x_{\mathcal{N}_c(j)} \in \mathbb{PP}_{\card{\mathcal{N}_c(j)}}-\frac{1}{2} & j\in\mathcal{J} \vspace{1ex}\\
		& x \in \left[-1/2, 1/2\right]^n
	\end{array}}
\end{equation}
is an equivalent optimization problem with two important differences.
The first is that the objective now penalizes closeness to $0$ rather
than to $1/2$.  The second is that check neighborhoods must be in the
``centered'' parity polytope.  The $d$-dimensional centered parity
polytope $\mathbb{PP}_{\card{\mathcal{N}_c(j)}}-\frac{1}{2}$ is
obtained by taking every point in $\mathbb{PP}_d$ and subtracting the
length-$d$ all $1/2$ vector.  For simplicity we subsequently refer to
this shifted object simply as the parity polytope, unless
disambiguation is required.

%% file: admmDecomposition.tex
In Section~\ref{sec.admmMsgPass}, we discuss the \gls{admm} algorithm, its application to
error-correction decoding and message passing interpretation.  The two
key subroutines: projection onto the parity polytope and projection
onto the probability simplex are discussed in
Sections~\ref{sec.parityPolyProject} and~\ref{sec.simpProj}, respectively.

%\gls{admm} is a large-scale decomposition technique for optimization.  
%It has recently become popular in machine learning and statistics as
%many optimization problems in these areas are
%decomposable~\cite{boyd_2011_admm}.  \gls{admm}'s macroscopic
%structure is similar to dual ascent.  It solves an equality
%constrained optimization by successively minimizing the problem's
%Lagrangian wherein dual variable estimates softly enforce equality
%constraints.  A solution is obtained by solving easier optimizations
%that can often be decomposed due to the absence of equality
%constraints.  Additionally, akin to the method of
%multipliers, \gls{admm} uses a regularized Lagrangian to smooth the
%dual problem and provide strong convergence guarantees.  Boyd et
%al.~\cite{boyd_2011_admm} describe \gls{admm} as ``an algorithm that
%is intended to blend the decomposability of dual ascent with the
%superior convergence properties of the method of multipliers''.  A
%distinct difference is that \gls{admm} is applied to problems whose
%primal variables can be partitioned into two sets from which the
%global objective is separable.  The two primal variable sets are
%related via a linear equality constraint, enforced via dual variable
%estimation as in dual ascent.

\subsection{ADMM Decomposition and Message Passing}
\label{sec.admmMsgPass}

The characteristic that, in a linear code, each component of the
codeword estimate $x$ (generally) participates in multiple check
constraints inhibits the decomposability of \gls{lp} decoding.
A small modification is therefore introduced in~\cite{draper_2013_admm_lp}
to apply \gls{admm} to \gls{lp} decoding .  We define an
auxiliary ``replica'' variable vector $z_j=x_{\mathcal{N}_c(j)}$ for
each check neighborhood.  By substituting into
(\ref{centeredLPDecoding}), we arrive at the following result, which fits the \gls{admm} template:
\begin{equation}
\label{admmLPDecoding}
{\begin{array}{r l l}
	\min & \gamma^\top x -\alpha\norm{x}{1}\\
	\text{subject to}& x\in [ -\frac{1}{2},\frac{1}{2} ]^n\\
	 & z_j \in \mathbb{PP}_{\card{\mathcal{N}_c(j)}}-\frac{1}{2} & j\in\mathcal{J} \\
	 & z_j=x_{\mathcal{N}_c(j)} & j\in\mathcal{J}.
	\end{array}}
\end{equation}

%As mentioned, t

The ADMM decomposition for (penalized) \gls{lp} decoding starts from
the $\ell_2$-regularized Lagrangian
\begin{equation*}
{\begin{array}{r l}
L_\mu \left( x, z, \lambda \right) = & \gamma^\top x - \alpha\norm{x}{1} + \sum\limits_{j\in \mathcal{J}}\lambda_j^\top \left(x_{\mathcal{N}_c(j)}-z_j\right)\\
& + \frac{\mu}{2}\sum\limits_{j\in \mathcal{J}}\norm{x_{\mathcal{N}_c(j)}-z_j}{2}^2
\end{array}}.
\end{equation*}
We use $z$ and $\lambda$ to refer to the $z_j$ and $\lambda_j$ in
aggregate.  The $\lambda_j$ are length-$\card{\mathcal{N}_c(j)}$ dual
variable vectors that enforce the $z_j=x_{\mathcal{N}_c(j)}$ equality
constraints.  The $\mu$ parameter is a positive number that determines
the degree of regularization.  While regularization does not change
the solution of the optimization problem, it accelerates algorithmic
convergence~\cite{boyd_2011_admm}.

\admmlp decoding alternates, in a round-robin manner, between
minimizing over codeword estimates $x$ and replicas $z$, followed by
an update of the dual variables $\lambda$.  Letting $\mathcal{X}$ and
$\mathcal{Z}$ represent the feasible sets of $x$ and $z$ (the dual
variables are unconstrained), each iteration takes the
form~\cite{draper_2013_admm_lp,draper_2014_penalized_lp}
\begin{equation}
{\begin{array}{r l l}
	x \gets & \arg\min_{x\in\mathcal{X}}  L_\mu \left( x, z, \lambda \right)\\
	
	z \gets & \arg\min_{z\in\mathcal{Z}}  L_\mu \left( x, z, \lambda \right)\\
	
	\lambda_j \gets & \lambda_j + \mu\left( x_{\mathcal{N}_c(j)}-z_j \right) & j\in\mathcal{J}.
\end{array}} \label{eq.roundRobin}
\end{equation}
The $x$ update can be decomposed into individual variable updates
since the solution to its optimization problem separates into distinct
calculations for each
variable~\cite{draper_2013_admm_lp,draper_2014_penalized_lp}.
Similarly, the $z$ update can be decomposed to update each $z_j$
individually.  The $\lambda$ update is already expressed in a
decomposed from.  When these update rules are fleshed out, $\mu$ can
be eliminated by reparameterizing $\gamma$, $\alpha$ and $\lambda$ by
a factor of $\mu$~\cite{wasson_2016_thesis}.  Therefore $\mu$ is not
included in the \admmlp decoding algorithm statement and the values of
$\gamma$, $\alpha$ and $\lambda$ have slightly different
parameterizations than in (\ref{eq.roundRobin}) going forward.

%%%%
% WOULD BE NICE TO NOTE THAT THE LAGRANGIAN IS A QUADRATIC FORM
%%%

The fact that the updates decompose means that the algorithm performs
a set of parallel variable updates followed by a set of parallel check
updates.  The result is a message-passing algorithm with a structure
similar to \gls{bp}.  Variable update $i$ is performed using
the \gls{llr} $\gamma_i$ and a length-$\card{\mathcal{N}_v(i)}$ vector
of messages from each of its neighboring checks, denoted
$m_{\mathcal{N}_v(i)\rightarrow i}$.  Check update $j$ is performed
using the dual variable vector $\lambda_j$ and a
length-$\card{\mathcal{N}_c(j)}$ vector of messages from each of the
neighboring variables.  The latter contains the current estimates
$x_{\mathcal{N}_c(j)}$ of neighboring variables.  The result of the
update of check $j$ is a new estimate of the associated dual variables
$\lambda_j$ as well as a length-$\card{\mathcal{N}_v(i)}$ message
vector whose components are sent to neighboring variables.  This
vector is denoted $m_{j\rightarrow\mathcal{N}_c(j)}$.
Alg.~\ref{decodingAlg} presents the \admmlp decoding algorithm in
full.  The notation $\prod_{\mathcal{A}}(\cdot)$ denotes Euclidean
projection onto the set $\mathcal{A}$.

\begin{figure}[t]
	%looks like this is how you make an ieee compliant algorithm box
	{\rule{0.5\textwidth}{0.4pt}}\vspace{-2.25ex}
	\AlgCaption{Given a \gls{llr} vector $\gamma\in\mathbb{R}^n$, calculate the $\ell_1$-penalized \gls{lp} decoding from (\ref{admmLPDecoding}) using \gls{admm}. The current estimate $x$ is returned upon termination.} 
	\vspace{0.3ex}
	\label{decodingAlg}
	{\rule{0.5\textwidth}{0.4pt}}

	\begin{algorithmic}[1]
		\State Initialize all $\lambda_j$'s and $m_{\mathcal{N}_v(i)\rightarrow i}$'s to 0.
		\Repeat
		\For{$i\in\mathcal{I}$} \Comment{\parbox{1in}{Variable Updates}}
			\State $t_i \gets 1^\top m_{\mathcal{N}_v(i)\rightarrow i} - \gamma_i $\label{varSumLine}
			\State $s_i \gets \left\{
			\begin{array}{ll}
			t_i + \alpha  & \mbox{if } t_i > 0 \\
			t_i & \mbox{if } t_i = 0 \\ %should this be put in??
			t_i - \alpha  & \mbox{if } t_i < 0
			\end{array}
			\right.$\label{varPenLine}
			\State $x_i \gets \prod\nolimits_{\left[-\frac{1}{2},\frac{1}{2}\right]} \left( \frac{s_i}{\left\vert{\mathcal{N}_v(i)}\right\vert}  \right)$\label{varNormLine}
		\EndFor %\vspace{1ex}
%		\Statex
		\For{$j\in\mathcal{J}$} \Comment{\parbox{1in}{Check Updates}}
			\State $v_j \gets x_{\mathcal{N}_c(j)} + \lambda_j$\label{chkStart}
			\State $z_j \gets \prod\nolimits_{\mathbb{PP}_{\left\vert{\mathcal{N}_c(j)}\right\vert} -\frac{1}{2} } \left( v_j \right)$ \label{chkProject}
			\State $\lambda_j \gets v_j - z_j$ \label{chkDualUpdate}
			\State $m_{j \rightarrow \mathcal{N}_c(j)} \gets 2z_j - v_j$\label{chkMsgUpLine}
		\EndFor
		
		\Until{termination}
	
	\end{algorithmic} \vspace{-2ex}
	{\rule{0.5\textwidth}{0.4pt}}
\end{figure}

\begin{figure}[t]
	\centering
	\includegraphics[width=0.3\textwidth]{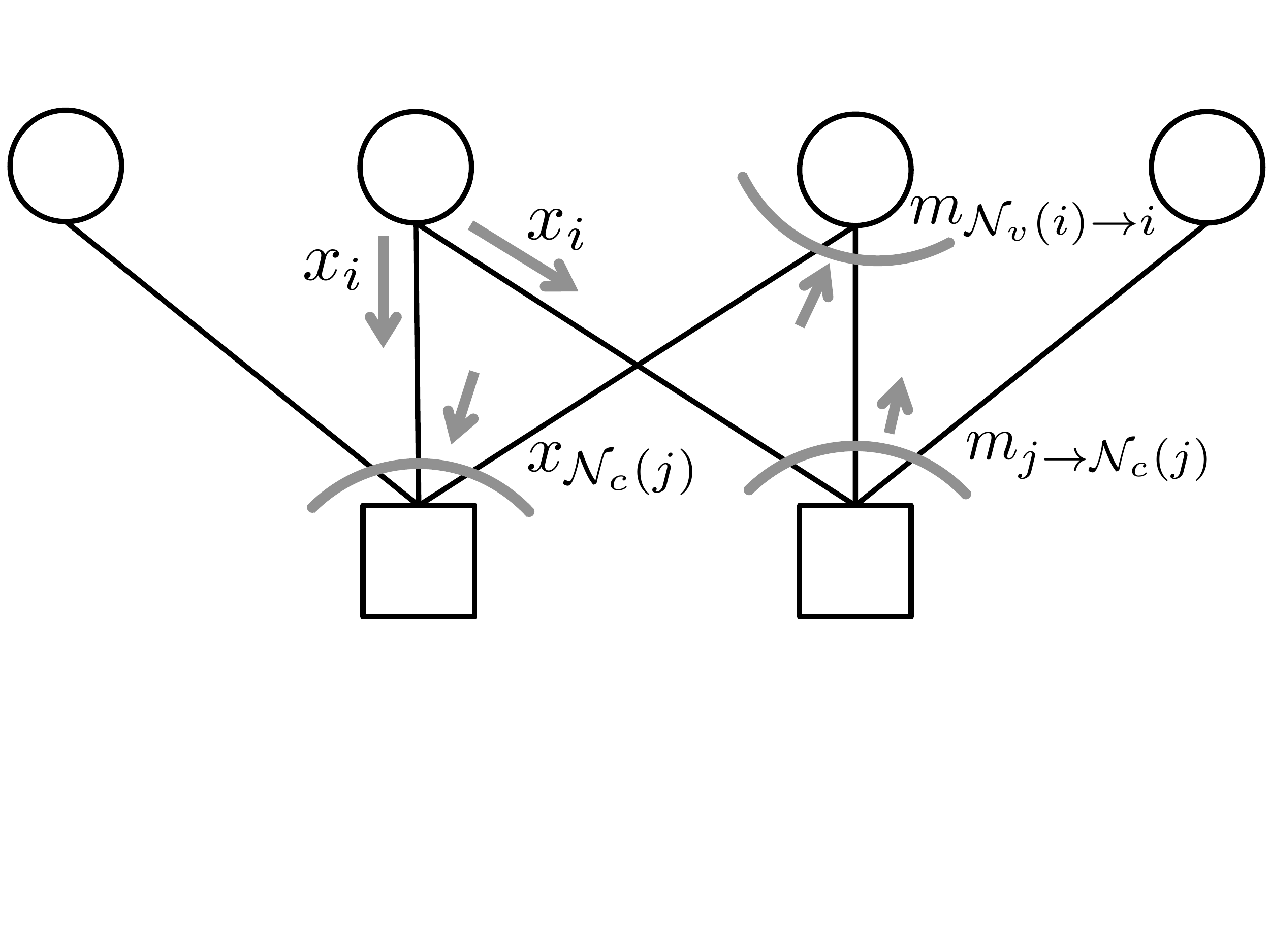}
	
	\vspace{0.1in}
	
	$H=\left(\begin{array}{cccc}
	1 & 1 & 1 & 0 \\
	0 & 1 & 1 & 1 \end{array}\right)$
	
	\FigCaption{A parity-check matrix and associated factor graph
	with messages labelled. Checks vertices are drawn as squares.
	}  \label{factorGraphFig}
\end{figure}

Viewing the variable/constraint structure of \admmlp
in~(\ref{eq.roundRobin}) using a factor graph, we observe that the
message-passing schedule followed in Alg.~\ref{decodingAlg} is the
standard flooding schedule of BP.  This view helps to highlight key
differences between \admmlp and \gls{bp}.  A factor graph is a
bipartite graph whose connectivity, when representing a linear code,
is specified by the code's parity-check
matrix~\cite{kschischang_2001_sum_product}.  For each codeword
variable there is a variable vertex.  For each parity check there is a
factor (or check) vertex.  Variable vertex $i$ and check vertex $j$
are connected if $H_{j,i}=1$.  The execution of \admmlp can be viewed
as variable updates taking place inside variable vertices and check
updates taking place inside check vertices.  Messages are passed
between variables and checks along the edges of the graph.  To
illustrate this interpretation, Fig.~\ref{factorGraphFig} depicts a
parity-check matrix and associated factor graph; edges are labeled
by \admmlp messages.

In Fig.~\ref{factorGraphFig} the $m_{\mathcal{N}_v(i)\rightarrow i}$
messages are composed of vector components from the
$m_{j \rightarrow \mathcal{N}_c(j)}$ messages.  The $\lambda_j$ do not
appear in the figure.  The vector of dual variables $\lambda_j$ is
used only within the $j^{\text{th}}$ check update.  Dual variable
vectors serve as internal check states that are not passed as
messages.  This is an important difference from \gls{bp} as the dual
variables play an important role in improving error-floor
performance~\cite{liu_2015_jumpLinear}.  A second difference is that
the $\card{\mathcal{N}_v(i)}$ outgoing messages from any variable $i$
are all equal, corresponding to the current estimate of $x_i$. This
differs from \gls{bp} where, in general, all outgoing messages from a
variable will differ.

We now examine the steps of Alg.~\ref{decodingAlg}.  On
line~\ref{varSumLine}, variable updates first sum incoming messages and
the negative \gls{llr} to form a variable estimate.  Incoming messages
tell the variable what value it should take on.  Next, on
line~\ref{varPenLine}, the penalization is applied.  A non-zero
penalty pushes each variable estimate in the direction of its current
belief.  Recall that this is done to discourage fractional solutions
(pseudocodewords).  When $\alpha$ is small, the effect of
penalization is reduced, making the algorithm closer to
(unpenalized) \gls{lp} decoding.  A slight difference in
Alg.~\ref{decodingAlg} from penalized \gls{lp} decoding's original
derivation in~\cite{draper_2014_penalized_lp} is that no penalty is
applied if $t_i=0$.  This modification is important in a fixed-point
implementation to avoid bias in codeword estimates.  On
line~\ref{varNormLine}, the penalized estimate is normalized by the
variable degree and projected onto the $[-\frac{1}{2},\frac{1}{2}]$
interval.  The resulting final estimate is then passed to neighboring
checks.  Roughly speaking, the variable estimate is the average of the
incoming messages. On line~\ref{chkStart}, the first step in the check update is to take
the vector of neighboring variable estimates and add the vector of
dual variables (the check state vector).  An updated vector of the
replica estimate is obtained by projecting the
addition result $v_j$ onto the parity polytope.  This is where the
parity polytope
constraints of \gls{lp} decoding are enforced.  Using the projection,
a new check state $\lambda_j$ and set of outgoing messages
$m_{j \rightarrow \mathcal{N}_c(j)}$ are calculated (possibly in
parallel) on lines \ref{chkDualUpdate}~and~\ref{chkMsgUpLine}.

%\tcr{(SCD: Mitch, I rewrote the following paragraph.  
%Your initial version is still in the .tex file (commented out).  I
%think this is an important paragraph that will take some massaging.
%Please check and edit. BTW, I hadn't really thought about this before,
%but why are the dual variables simply the difference between $z_j$ and
%$v_j$?  Is it related to the fact that when we project $v_j$ onto the
%simplex we adjust each (active) component by the same amount because
%we are moving along a 45-degree line coming out of the origin?  This
%should relate to the KKT question I put on last year's final.  Let's
%see if we can formulate a one-sentence explanation.)}

%To stark from mith: I am more sure of the first point below since we
%talked on Nov 7.

We think of the dual variable estimates as affecting algorithmic
progression in two major ways. First, $\lambda_j$ acts as a momentum
term on line~\ref{chkStart}. It brings $v_j$ closer to the previous
value of $v_j$, ensuring that $z_j$ does not evolve too erratically.
Second, according to the pricing interpretation of duality, the
$\lambda_j$ specifies the cost of breaking the equality constraint
$z_j=x_{\mathcal{N}_c(j)}$.  We can see this effect more clearly if
line~\ref{chkMsgUpLine} is rewritten as
$m_{j \rightarrow \mathcal{N}_c(j)} \gets z_j - (v_j - z_j) = z_j
- \lambda_j$.  Since the new $\lambda_j$ value is the mismatch between
$v_j$ and $z_j$, line~\ref{chkMsgUpLine} compensates this mismatch by
including it in the outgoing messages.  At convergence, the
$\lambda_j$ subtracted off here is canceled by the $\lambda_j$ added
in to compute $v_j$ on line~\ref{chkStart}.

Note that a termination condition is not specified in
Alg.~\ref{decodingAlg}.  While algorithmic convergence can be used as
the stopping criterion in floating point~\cite{draper_2013_admm_lp},
it may not be possible to obtain convergence in fixed point to an
arbitrary precision.  Thus, in our implementation, we impose a fixed
number of iterations, but can also terminate early if rounding the
current codeword estimate produces a codeword.

%\tcr{(SCD: So the latter would be useful, e.g., if we converge to a 
%pseudo-codeword, so checking whether rounded estimate is a codeword
%wouldn't suffice?)}

%Because of its message-passing structure, the complexity of \admmlp
%scales linearly in blocklength~\cite{draper_2013_admm_lp}.  \admmlp
%decoding thus well suited to \gls{ldpc}
%codes~\cite{gallager_1962_ldpc}.  As for \gls{bp}, the small check
%degrees of \gls{ldpc} codes provide a computational advantage.

While their message-passing schedules are the same, we have already
observed two differences between \admmlp and \gls{bp}: the existence of 
dual variables that form the check states, and the fact that all
outgoing messages from a variable node are identical.  To this point, a
third significant difference has been abstracted: the
computational primitive of the check update, which is the Euclidean
projection onto the parity polytope.  A discussion of how to implement
this projection efficiently in hardware is the topic of the next
section.

%% file: parityPolytope.tex
\subsection{Parity Polytope Projection}
\label{sec.parityPolyProject}

%%%%%%%%%%%%%%%%%%%%%%%%%%%%%%%%%%%%%%%%%%%%%%%%%%%%%%%%%%%
%this figure needs to be moved around in the tex doucment to get it on the right page.
\begin{figure}[t]
	\centering \subfloat[Identify active
          facet]{\includegraphics[width=.15\textwidth]{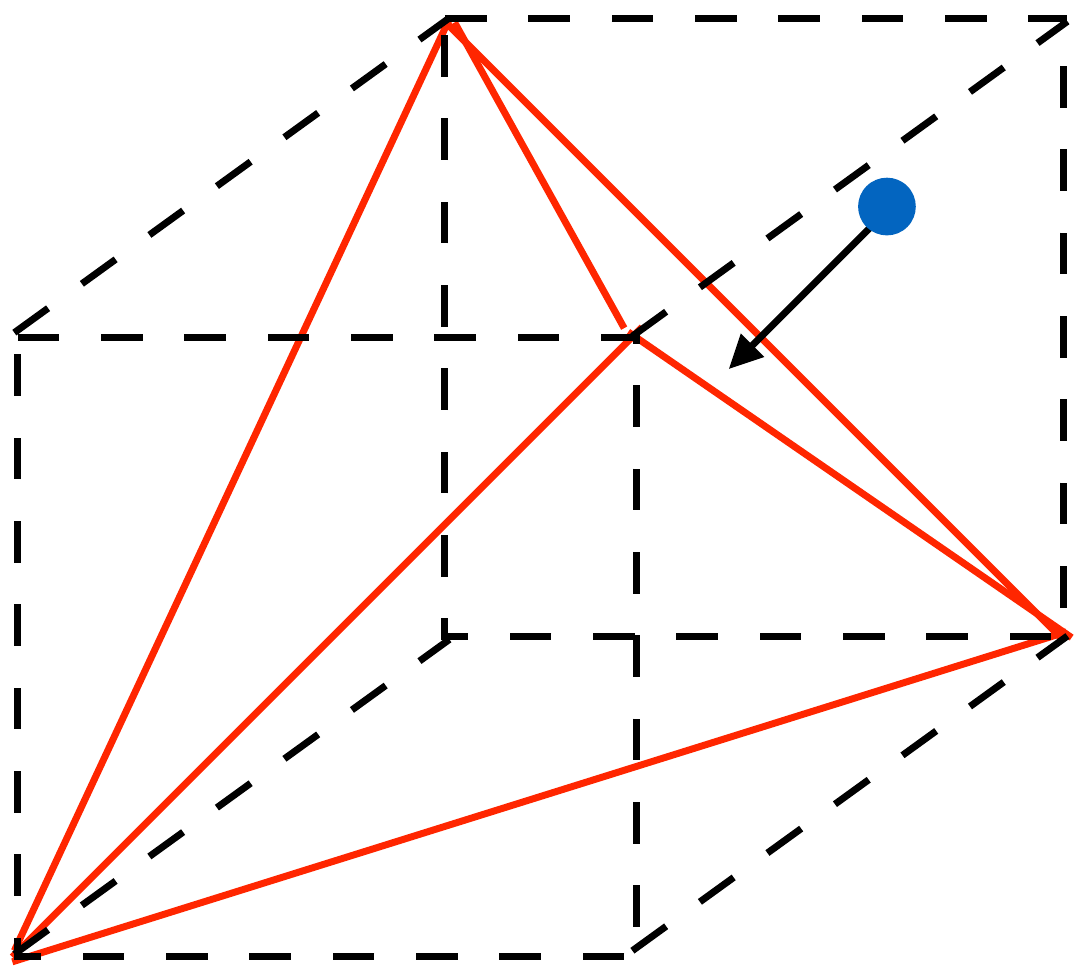}\label{ppProjectA}}
        \hfill \subfloat[Transform
          problem]{\includegraphics[width=.15\textwidth]{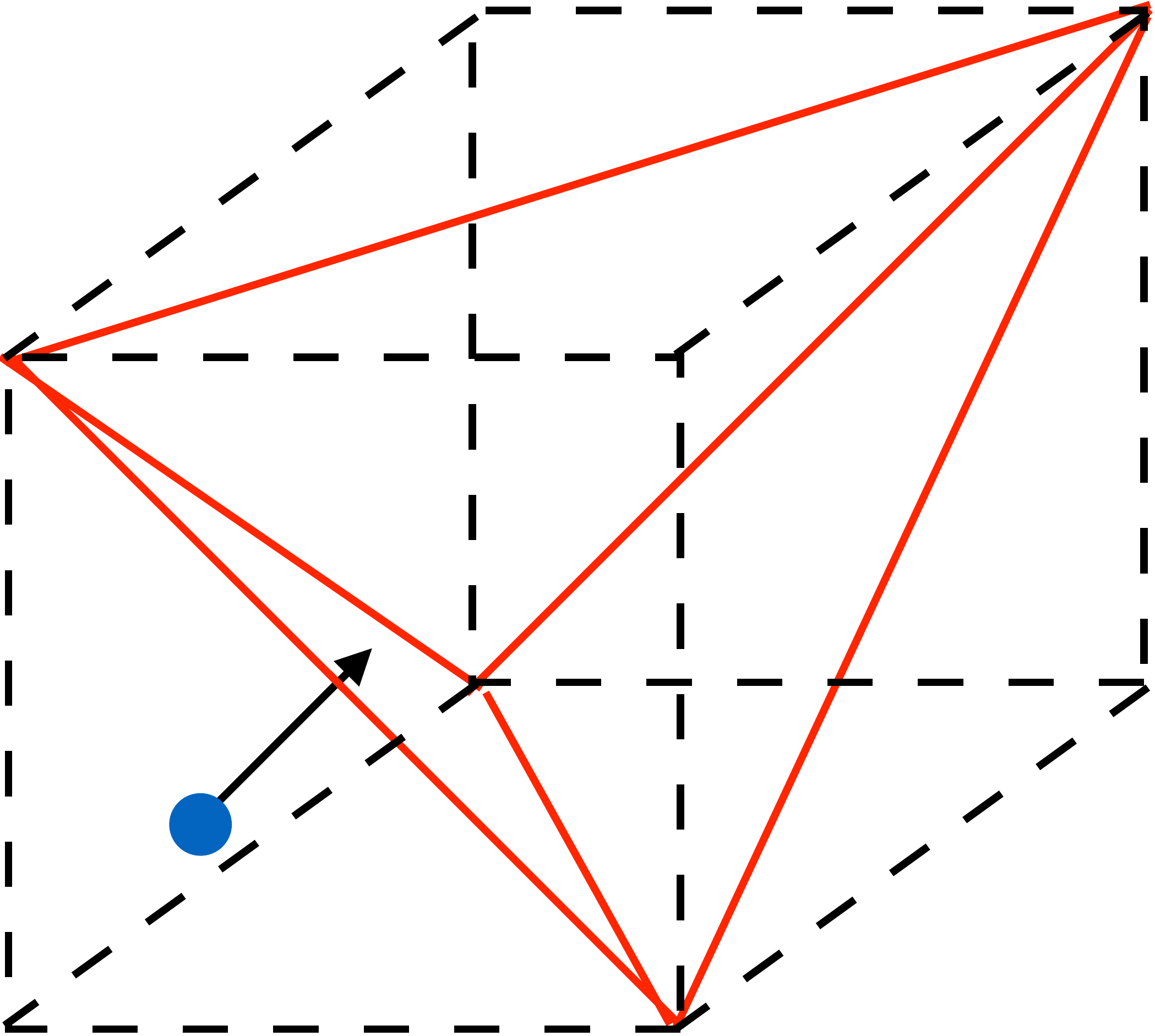}\label{ppProjectB}}
        \hfill \subfloat[Simplex projection]{\includegraphics[width=.15\textwidth]{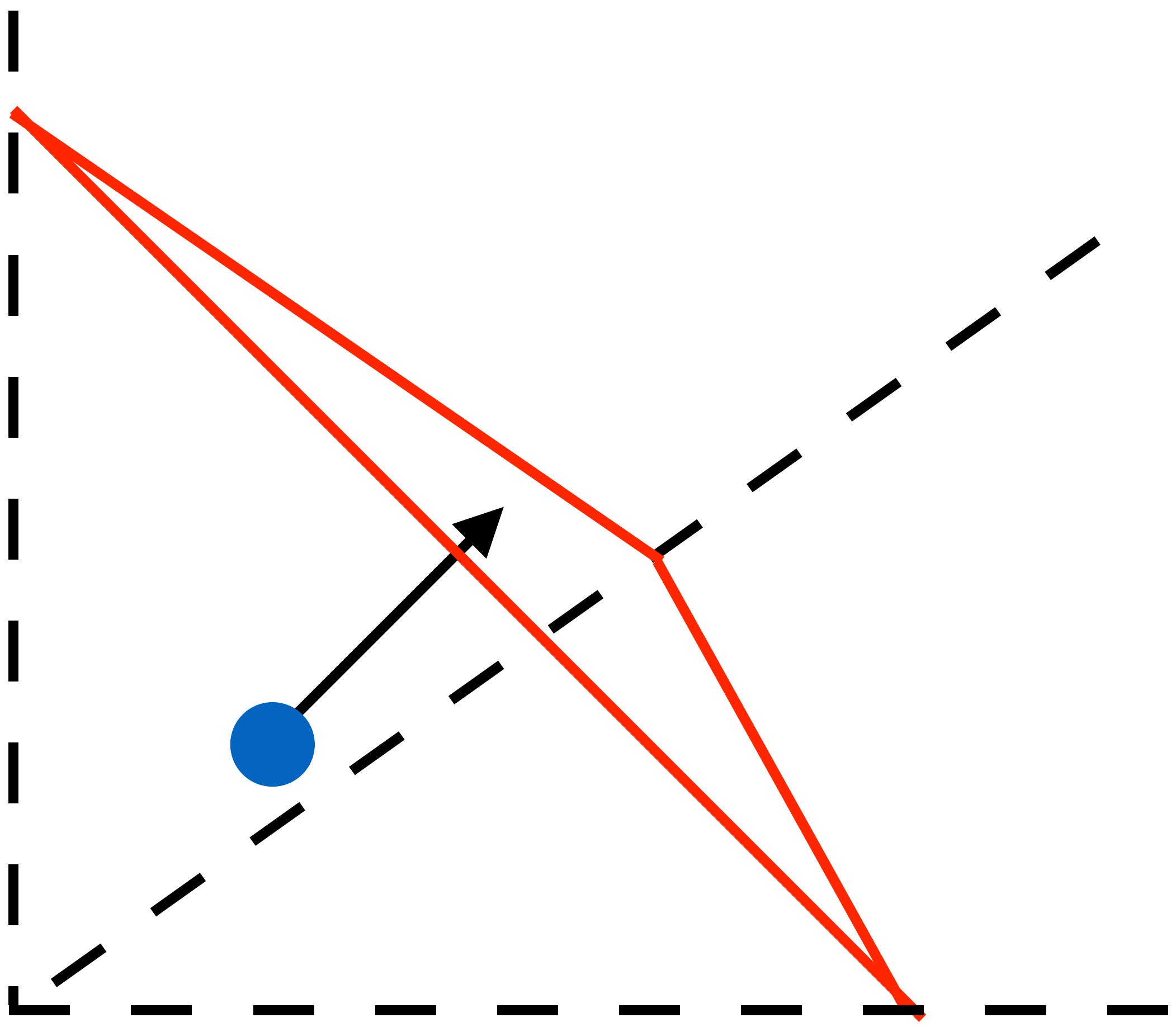}\label{ppProjectC}}
        \hfill \FigCaption{Projection onto the parity polytope
          $\mathbb{PP}_3 \! - \! \frac{1}{2}$: identify active facet,
            transform into canonical coordinate system, project
            onto probability simplex.}
	\label{ppProject}
\end{figure}
%%%%%%%%%%%%%%%%%%%%%%%%%%%%%%%%%%%%%%%%%%%%%%%%%%%%%%%%%%%

Euclidean projection of a vector $v$ onto the $d$-dimensional parity
polytope is specified by the quadratic program
\begin{equation}
	\prod\nolimits_{\mathbb{PP}_d-\frac{1}{2}}(v) = \arg\min_{w\in\mathbb{PP}_d-\frac{1}{2}} \norm{w-v}{2}^2.
\end{equation}
Projection onto the centered $(\mathbb{PP}_d-\frac{1}{2})$ and
non-centered $(\mathbb{PP}_d)$ parity polytope are  similar
operations related as
\begin{equation*}
\prod\nolimits_{\mathbb{PP}_d-\frac{1}{2}}(v) = \prod\nolimits_{\mathbb{PP}_d}\left(v+\frac{1}{2}\right)-\frac{1}{2}.
\end{equation*} 

Barman et al. began the investigation into efficient projection onto
the parity polytope~\cite{barman_2011,draper_2013_admm_lp}.  These
researchers established a ``two-slice'' representation of the polytope
and exploited rotational symmetry to sort the components of $v$ into a
canonical coordinate system for projection (and subsequent
de-sort). However, their algorithm is not well-suited for hardware due
to its iterative nature and complexity of the sorting
procedure. X.~Zhang and
Siegel~\cite{siegel_2013_projection_lp,zhang_2012_phd} improved the
method by removing the sort and de-sort operations through efficient
identification of the violated cut from~(\ref{ppCutDef}).
Unfortunately, as with the first approaches, the method remains
intensively iterative.  In parallel
to~\cite{siegel_2013_projection_lp}, G.~Zhang et al.~made the
connection to projection onto the probability
simplex~\cite{kleijn_2013_pp_projection} which provides clean
geometric intuition.  In~\cite{wasson_2015_pp} Wasson and Draper
combined the advances of~\cite{kleijn_2013_pp_projection,
  siegel_2013_projection_lp} to create the hardware-compatible method
of projection we now describe.

First, the violated cut from (\ref{ppCutDef}) is identified, revealing
the active facet.  The identified cut defines a similarity transform
used to reorient the problem into a canonical orientation.  The
problem is thereby reduced to projection onto the (centered)
probability simplex.  After projecting onto the simplex, the
similarity transform is inverted to yield the projection onto the
parity polytope.  This high-level description is depicted in
Fig.~\ref{ppProject}.  The algorithm described in Alg.~\ref{ppProjAlg}
was slightly modified in~\cite{wasson_2016_thesis} from the algorithm
presented in~\cite{wasson_2015_pp} to project onto the centered parity
polytope.  The algorithm has a straightforward, non-iterative,
execution path whose steps can largely be parallelized.  This,
combined with simple intuition, makes Alg.~\ref{ppProjAlg} an
excellent candidate for hardware adoption.

\begin{figure}[t]
	%looks like this is how you make an ieee compliant algorithm box
	%might figure out later how to do algorithm numbering.
	%for now, just reference it as a figure.
	{\rule{0.5\textwidth}{0.4pt}} \vspace{-4.5ex}
	\AlgCaption{Given a vector $v\in\mathbb{R}^d$, calculate its Euclidean projection onto the parity polytope. $w=\prod\nolimits_{\mathbb{PP}_d-\frac{1}{2}}(v)$ is returned from the method.} 
	\vspace{0.3ex}
	\label{ppProjAlg}
	{\rule{0.5\textwidth}{0.4pt}}
	
	\begin{algorithmic}[1]		
		\For{$i\in [d]$} \Comment{\parbox{1.175in}{Facet Identification}}\label{facetStart}
		\State $
		f_i \gets
		\begin{cases}
		1& \text{if } v_i \geq 0\\
		0           & \text{otherwise}
		\end{cases}
		$\label{fConst}
		\EndFor
		
		\If{$1^\top f$ is even \label{hammWt}}
			\State $i^\ast \gets \arg\min_{i\in[d]} |v_i|$ \label{argMin}
			\State $f_{i^\ast} \gets 1- f_{i^\ast}$\label{fChange}
		\EndIf\label{facetEnd}
		
		%\Statex
		
		\For{$i\in [d]$} \Comment{\parbox{1.175in}{Similarity Transform}}\label{simTran1Start}
			\State $\tilde{v}_i \gets v_i {\left(-1\right)}^{f_i}$
			%got rid of conditional statement
			\iffalse
			\begin{cases}
			-v_i& \text{if } f_i = 1\\
			v_i           & \text{otherwise}
			\end{cases}
			\fi
			\label{flipPosCoor}
		\EndFor\label{simTran1End}
		
		%\Statex
		
		\State $\tilde{u} \gets \prod\nolimits_{\mathbb{S}_d -\frac{1}{2} }\left( \tilde{v} \right)$ \Comment{\parbox{1.175in}{Simplex Projection}}
		
		%\Statex
		
		\For{$i\in [d]$} \Comment{\parbox{1.175in}{Similarity Transform}}
			\State $u_i \gets \tilde{u}_i{\left(-1\right)}^{f_i}$
			\iffalse 
			\begin{cases}
			-\tilde{u}_i& \text{if } f_i = 1\\
			\tilde{u}_i          & \text{otherwise}
			\end{cases} 
			\fi
			 \label{simTransInvert}
		\EndFor \label{endShellProj}
		
		%\Statex
		
	%	\State $a \gets 1^\top \prod\nolimits_{\left[-\frac{1}{2},\frac{1}{2} \right]^d}\left( \tilde{v} \right)$
		\If{$ 1^\top \prod\nolimits_{\left[-\frac{1}{2},\frac{1}{2} \right]^d}\left( \tilde{v} \right)  \geq 1 - \frac{d}{2}$}\label{test}\Comment{\parbox{1in}{Membership Test}}
			\State $w \gets \prod\nolimits_{\left[-\frac{1}{2},\frac{1}{2} \right]^d}\left( v \right)$
		\Else
			
			\State $w \gets u$ \label{caseOutside}
		\EndIf \label{endMembershipTest}
		
	\end{algorithmic}\vspace{-2ex}
	{\rule{0.5\textwidth}{0.4pt}}
\end{figure}

In Alg.~\ref{ppProjAlg}, lines~\ref{facetStart}--\ref{facetEnd} form
the facet identification portion of the projection algorithm.  The
objective here is to identify the vertex cut from (\ref{ppCutDef})
that is violated (if one is violated).  This amounts to finding the
closest odd-weight vertex of the unit
hypercube~\cite{siegel_2012_cut_search}.  First, the closest vertex of
the hypercube is found and stored in the binary vector $f$.
%(One could alternately compute $1-f$ which is easily obtained as the sign bit of the fixed-point representation of $v$.)  
For the non-centered case considered in Alg.~\ref{ppProjAlg}, the
actual vertex is $f - 0.5$.  If the Hamming weight of $f$ is odd, then
the closest vertex violates the parity constraint (i.e., it is {\em
  not} a codeword of the single parity-check code) and we have
identified the violated cut.  On the other hand, if the Hamming weight
of $f$ is even, the nearest vertex does not violate the parity
constraint.  The Hamming weight computation is performed on
line~\ref{hammWt}.  To find the violated cut, we perturb the $f$
vector in one coordinate.  The coordinate to perturb corresponds to
the $v_i$ that is closest to the midpoint of the unit interval.  This
coordinate is identified on line~\ref{argMin} and $f$ is perturbed
accordingly to make it of odd weight on line~\ref{fChange}.

%We note that the test $v_i \geq 0$ on line~\ref{fConst} can be changed
%to $v_i > 0$ without affecting the output of the algorithm.  There are
%two cases where the change to $v_i > 0$ might impact the projection.
%First, if a single $v_i$ component is identically $0$, the
%corresponding $f_i$ will be the changed component on
%line~\ref{fChange}.  Second, if multiple $v_i$ components are
%identically $0$, this means that there are multiple closest odd-weight
%vertices.  As we will explain shortly, it does not matter which
%closest odd-weight vertex is chosen in the latter case.

Once the possibly violated cut is known, a similarity transform
applied on line~\ref{simTran1Start} transforms $v$ to $\tilde{v}$.
This aligns the identified cut with the (centered) probability
simplex.  This transformation is illustrated in Fig.~\ref{ppProject}
where $v$ is the dot in Fig.~\ref{ppProjectA}, $\tilde{v}$ is the dot
in Fig.~\ref{ppProjectB}, and $f = [1 \, 1 \, 1]$ (or $f = [0.5 \, 0.5
  \, 0.5]$). The transformed point $\tilde{v}$ is then projected onto
the (centered) probability simplex, as illustrated in
Fig.~\ref{ppProjectC}.  After projection, the similarity transform is
inverted on line~\ref{simTransInvert}.  The similarity transform is
self-inverting.

The execution path up through line~\ref{endShellProj} of
Alg.~\ref{ppProjAlg} produces a projection onto the boundary or
``shell'' of the parity polytope.  Through these steps, a point already
inside the parity polytope, instead of being left unperturbed, would
be projected onto the cut corresponding to the closest odd-weight
vertex.  To avoid this, we test for parity polytope membership on line~\ref{test}.  

We now describe a test for parity polytope membership.  If the vector
being tested is in the unit hypercube, we need only to check the
previously identified cut~\cite{siegel_2012_cut_search}.
Line~\ref{test}, originally given in~\cite{zhang_2012_phd}, tests the
hypercube projection of $v$ against the identified cut.  This is done
by taking the hypercube projection of $\tilde{v}$ and checking on
which side of the (centered) probability simplex it lies.  If the
hypercube projection of $v$ is in the parity polytope, then this must
be the parity polytope projection of $v$ since the parity polytope is
a subset of the hypercube.  A point already in the parity polytope
will be left unperturbed.

%% file: simplex.tex
\subsection{Simplex Projection}
\label{sec.simpProj}

\begin{figure}[t]
	%looks like this is how you make an ieee compliant algorithm box
	%might figure out later how to do algorithm numbering.
	%for now, just reference it as a figure.
	{\rule{0.5\textwidth}{0.4pt}}\vspace{-2ex}
	\AlgCaption{Given a vector $v\in\mathbb{R}^d$, calculate its Euclidean projection onto the centered probability simplex. $w=\prod\nolimits_{\mathbb{S}_d-\frac{1}{2}}(v)$ is returned from the method.} 
	\vspace{0.03in}
	\label{simpProjAlg}
	{\rule{0.5\textwidth}{0.4pt}}
	
	\begin{algorithmic}[1]		
		\State $\rho \gets \text{descendingSort}\left(v\right)$\label{sortLine}
		%\Statex
		\For{$i\in [d]$}  \Comment{\parbox{1.4in}{Calculate Possible Shifts}} \label{computeShiftStart}
			\State $u_i \gets \frac{1}{i}\left(\sum\limits_{j=1}^{i} \rho_j - 1\right)$
		\EndFor
		%\Statex
		\State $i^\ast \gets \max\left\{ i \in [d] : \rho_i > u_i \right\}$ \Comment{\parbox{1.3in}{Choose Shift}}\label{chooseShift}
		%\Statex
		\For{$i\in [d]$} \Comment{\parbox{1.3in}{Perform Shift and Clip}}
			\State $w_i \gets \max\left\{ v_i-u_{i^\ast} -\frac{1}{2},-\frac{1}{2} \right\}$ \label{implementShift}
		\EndFor
		
	\end{algorithmic}\vspace{-2ex}
	{\rule{0.5\textwidth}{0.4pt}}
\end{figure}

%%%%%%%%%%%%%%%%%%%%%%%%%%%%%%%%%%%%%%%%%%%%%%%%%%%%%%%%%%%
%this figure needs to be moved around in the tex doucment to get it on the right page.
%\begin{figure*}[ht]
%	\centering
%	\subfloat[Desired simplex projection.]{\includegraphics[width=.3\textwidth]{simpProjA}} \hfill
%	\subfloat[Shift along all $1$'s vector.]{\includegraphics[width=.3\textwidth]{simpProjB}} \hfill
%	\subfloat[Clip components if necessary.]{\includegraphics[width=.3\textwidth]{simpProjC}} \hfill
%	\FigCaption{Projection onto the centered probability simplex $\mathbb{S}_2-\frac{1}{2}$.}
%	\label{simpProject}
%\end{figure*}
%%%%%%%%%%%%%%%%%%%%%%%%%%%%%%%%%%%%%%%%%%%%%%%%%%%%%%%%%%%

We now consider the final important algorithm: Alg.~\ref{simpProjAlg},
projection onto the centered probability simplex.  The centered
probability simplex $\mathbb{S}_d\!-\!\frac{1}{2}$ is defined by
subtracting the all-$1/2$ vector from the probability simplex
$\mathbb{S}_d= \left\{ v\in\mathbb{R}^d : 1^\top v = 1, \text{ } v_i
\geq0 \text{ }\forall \text{ }i\in[d] \right\}$.  Projection of $v \in
\mathbb{R}^d$ onto the centered probability simplex is a quadratic
program.
%\begin{equation}
%\prod\nolimits_{\mathbb{S}_d-\frac{1}{2}}(v) = %\arg\min_{w\in\mathbb{S}_d-\frac{1}{2}} \norm{w-v}{2}^2.
%\end{equation} 
Additionally, projection onto the centered and non-centered
probability simplexes are related in the same manner as for the centered and
non-centered parity polytope.

%\begin{equation*}
%\prod\nolimits_{\mathbb{S}_d-\frac{1}{2}}(v) = %\prod\nolimits_{\mathbb{S}_d}\left(v+\frac{1}{2}\right)-\frac{1}{2}.
%\end{equation*}

%Two methods for projecting onto the probability simplex were studied
%by Duchi et al.~\cite{duchi_2008_simplex_projection}.  While very
%efficient, the second method presented in
%\cite{duchi_2008_simplex_projection} is highly iterative with many
%different execution paths.  For this reason, it is not suitable for a
%hardware-based implementation.  

Algorithm~\ref{simpProjAlg} presents a simplex projection method
from~\cite{duchi_2008_simplex_projection}, modified to project onto
the centered simplex.  Indeed, computing the projection is a rather
straightforward optimization easily solved through analyzing the
\gls{kkt}
conditions~\cite{duchi_2008_simplex_projection,wasson_2016_thesis}.
The \gls{kkt} conditions tell us that the projection is obtained by
shifting $v$ along the all-$1$s vector, clipping components that fall
below 0, and ensuring non-clipped components sum to
$1$. The shift along the all-ones vector results from the fact that
the all-$1$s vector is orthogonal to the simplex, as shown in
Fig.~\ref{ppProjectC}.  The clipped components are the most
negative components of $v$, therefore the $d$ possible shifts are
computed from a sorted version of $v$.  A smart way to identify the
best common shift is developed in~\cite{ss_2006_simplex} and used
herein.  The magnitude of the common shift is computed on
lines~\ref{computeShiftStart}-\ref{chooseShift} of
Alg.~\ref{simpProjAlg}.  The shift and clip are implemented on
line~\ref{implementShift}.

%The heart of the simplex projection problem is in determining the
%common shift value and which components get clipped.  \gls{kkt}
%analysis shows that the common shift value is the value which makes
%non-clipped components sum to $1$.  Additionally, in
%\cite{ss_2006_simplex}, Shalev-Shwartz and Singer proved that the
%clipped components correspond to the most negative components of $v$.
%The projection algorithm in Fig~\ref{simpProjAlg} uses this
%information by calculating the $d$ possible shift values from a sorted
%version of the input vector.  Each shift value could be tried to see
%which gives the smallest objective function value, but Shalev-Shwartz
%and Singer provided a better way to identify the
%shift~\cite{ss_2006_simplex}.  Their method is used on
%line~\ref{chooseShift} to choose the optimal shift value.

%% file: decoderArch.tex
%\section{Architectures}
%\begin{figure*}
%	\centering
%	\includegraphics[width=0.89\textwidth]{arch_MM}
%	\caption{Partially-parallel decoder architecture.}
%	\label{decoderArchitecture}
%\end{figure*}
\begin{figure}[t]
	\centering
	%play with width to match space constraints
	\includegraphics[width=0.49\textwidth]{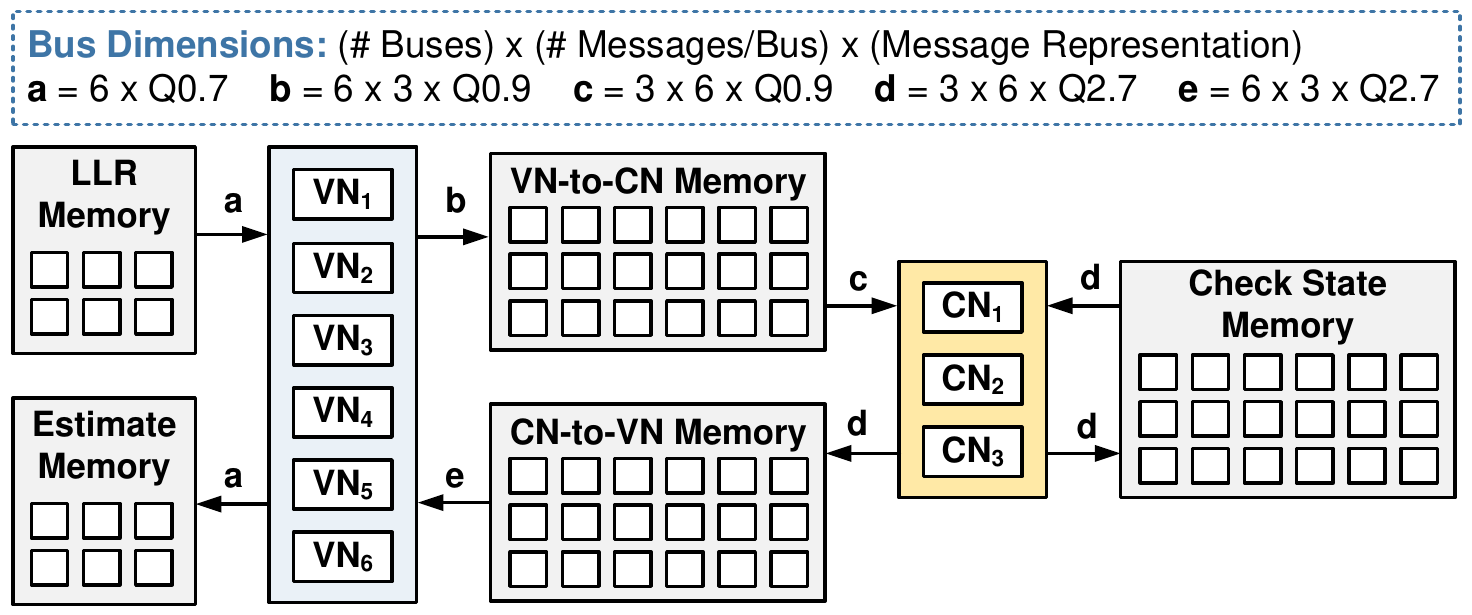}
	\caption{Partially-parallel architecture for a ($3$, $6$)-regular \qcldpc code.}
	\label{decoderArchitecture}
\end{figure}

In this section, we build upon well-known hardware architectures for
message-passing decoders in the design of a hardware-based \admmlp
decoder implementation.  We modify the arithmetic kernels in the check
and variable processing nodes per the operations outlined in
Alg.~\ref{decodingAlg}.

%In the previous section, we showed that \admmlp decoders implement a
%message-passing schedule similar to \gls{bp} decoders, one that is
%based on the implied check and variable node connectivity structure
%defined by a binary parity-check matrix.  This similarity allows us to
%build upon well-known hardware architectures used for \gls{bp}
%decoders in the design of a hardware-based \admmlp decoder
%implementation.  We modify the arithmetic kernels in the check and
%variable processing nodes as per the operations outlined in
%Alg.~\ref{decodingAlg}.

\subsection{Decoder}

A central challenge in implementing hardware-based decoders is the
scalability of the message-passing network.  We restrict ourselves to \gls{qc} codes~\cite{kou_2001_qc,fossorier_2004_qc} and
a partially-parallel architecture~\cite{Hocevar2004}.  This
simplifies message routing and memory interfacing.  We
implement the message-passing network with regularly-distributed
on-chip \gls{fpga} block RAMs.  Figure~\ref{decoderArchitecture}
presents an overview of our partially-parallel \qcldpc decoder
architecture for the special case of a $(3,6)$-regular QC-LPDC code.
The architecture is comprised of multiple memory types to store input
\gls{llr}s, intermediate messages, and output codewords, as well as
pipelined \gls{cn} and \gls{vn} processing units that perform the
arithmetic operations of Alg.~\ref{decodingAlg}.

\qcldpc codes are defined by a parity-check matrix formed by
tilings of $p\times p$ circulant matrices.  Each tile of a \gls{qc}
parity-check matrix can be either the all-zeros matrix or some
addition of shifted-identity matrices.  The tilings naturally divide
the parity-check matrix into $s=\frac{n}{p}$ ``macro''-columns and
$r=\frac{m}{p}$ macro-rows.  Inside a given macro-row (column), the
required message locations for a check (variable) computation are the
locations for the previous check (variable) plus 1 modulo $p$.  This
rich class of codes is popular in hardware implementations, appearing
in standards such as IEEE 802.11ad~\cite{IEEE-80211ad_rev}.

The first step our decoder executes is to load \gls{llr}s into memory.
We instantiate $s$ memories, each of depth $p$, to store the
\gls{llr}s.  Each memory is read in parallel to feed \gls{llr}s into
$s$ pipelined \gls{vn}s.  The \gls{vn}s also receive messages from a
\gls{ctv} message memory, to be discussed later.  At the output of the
\gls{vn}s, the current variable estimates $x_i$ are written in
parallel into $s$ estimate memories, to be read from upon termination.
Variable estimates are also written into \gls{vtc} message
memories.  There is a \gls{vtc} message memory for each
shifted-identity matrix in the specification of the parity-check
matrix.  These memories are addressed using their corresponding shift
number to ensure the messages are passed to the proper \gls{cn}.

Next, $r$ pipelined \gls{cn}s read their required messages in parallel
from the \gls{vtc} message memory.  In addition, check states are read
from check state memories, which are instantiated in the same manner
as the \gls{vtc} message memories.  However, address shifting is not
required since these memories are only accessed by \gls{cn}s.  
When a \gls{cn} computation completes, the new check
states are written into check state memories and the messages are
written into \gls{ctv} message memories.  The \gls{ctv} message
memories are structured in the same manner as the \gls{vtc} memories,
with write operations using cyclic shift information.  The process
repeats until the maximum number of iterations is exceeded, or some
early termination condition is satisfied.

We find our current implementation of the \admmlp decoder to be
sensitive to fixed-point quantization.  Min-Sum decoders can be
implemented with 5- or 6-bit message widths while suffering minimal
degradation in error-rate performance compared to
floating-point~\cite{Park2014}.  \admmlp requires larger bit-widths.
We believe the higher precision is required because the result of the
projection operation that \gls{cn}s perform must be quantized.  The
quantization results in a loss of precision and a corresponding
deterioration of message resolution.

We now discuss the logic that underlies the choices we made in
selecting fixed-point representations.  We first note that a change in
the assignment of bits between integer and fraction parts of
fixed-point \gls{llr}s amounts to a linear scaling of the \gls{lp}
objective.  However, any scaling of the objective in an LP (i.e., of
$\gamma$ in~(\ref{centeredLPDecoding})) does not change the solution
of the LP.  This provides some flexibility in choosing the fixed-point
representation of the \gls{llr}s.  Next, we note that each message
passed to a \gls{vn} can be thought of as either trying to overcome
the channel information or as trying to reinforce it.  Thus, we
allocate any extra bit-width to the integer part of a \gls{ctv}
message.  This provides the dynamic range required to override channel
\gls{llr}s.  In contrast, extra bit-width allocated to \gls{vtc}
messages should be in the fractional part.  An increase in the number
of fraction bits mitigates the effect of the inexact (due to finite
precision) normalization by $|\mathcal{N}_{v\left( i \right)}|$ in the
\gls{vn}s.

Based on this intuition, we select fixed-point message representations
to retain as much channel information as possible.  We first consider
the bit-width of \gls{llr}s and the estimate outputs.  Respectively,
these correspond to the decoder's input and output message widths.
Next, we consider how many additional bits \gls{vtc} and \gls{ctv}
messages will receive.  \gls{vtc} messages, as well as the estimates,
lie in the centered unit hypercube.  Therefore, these messages receive
one sign bit and no integer bits.  Next, we give \gls{llr}s one sign
bit, zero integer bits, and allocate the remainder to fraction bits.
This ensures that all channel information is visible in the estimates
and the \gls{vtc} messages.  Experimentation shows that this
fixed-point \gls{llr} representation provides the best error-rate
performance~\cite{wasson_2016_thesis}.  The \gls{ctv} messages are
given one sign bit and the same number of fraction bits as the LLRs.
This is done so that the summation in the \gls{vn} computation
produces an output that does not have any constant bits for some given
\gls{llr}.  Finally, the check states are given the same
representation as the \gls{ctv} messages because they are computed in
a similar manner.

%% file: variableNode.tex
\subsection{Variable Node}

\begin{figure}[t]
	\centering
	\includegraphics[width=0.48\textwidth]{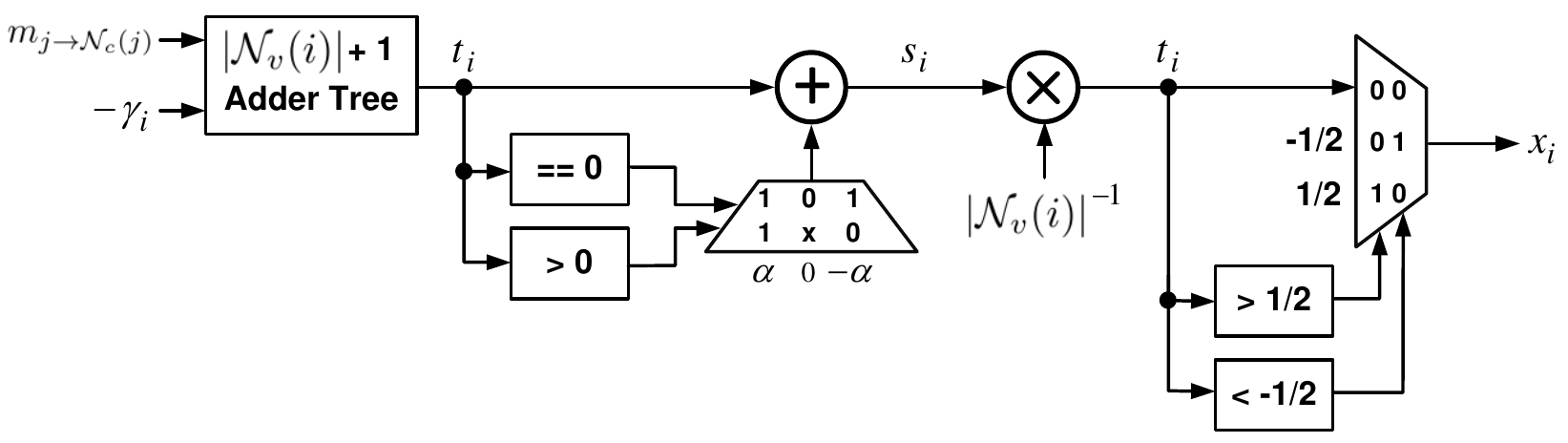}
	\FigCaption{Variable node compute module.}
	\label{vnArch}
\end{figure}

A \gls{vn} executes lines \ref{varSumLine}--\ref{varNormLine} of
Alg.~\ref{decodingAlg}.  An implementation schematic for a \gls{vn},
labeled with variables from Alg.~\ref{decodingAlg}, is depicted in
Fig.~\ref{vnArch}.  A \gls{vn} starts by summing the incoming messages
$m_{\mathcal{N}_v(i)\rightarrow i}$ and subtracting the \gls{llr}
$\gamma_i$.
%In the proposed architecture \gls{llr}s and \gls{ctv}
%messages have the same number of fraction bits.  Therefore their
%fixed-point representations can be added without modification.  
This
large addition operation is performed using a pipelined adder tree
with $\ceil{\log_2\left( \card{\mathcal{N}_v(i)} + 1 \right)}$ adder
stages with pipelining in between.  The output of the adder tree is
provided an additional $\ceil{\log_2\left( \card{\mathcal{N}_v(i)} + 1
  \right)}$ integer bits to prevent overflow.

To implement penalization, the \gls{vn} checks to see if the adder
tree result $t_i$ is greater than, equal to, or less than 0.  Using
this information, two multiplexers then choose to add $\alpha$, $0$,
or $-\alpha$ to the adder tree output.  An integer bit is added to the
fixed-point representation to avoid overflow.

The next step in the \gls{vn} is to normalize the penalized sum $s_i$.
Division is generally an expensive operation to perform,
but variable degrees are constant for a given code.  Therefore,
division by $\card{\mathcal{N}_v(i)}$ can be performed by finding its
reciprocal during synthesis and executing the normalization with a
multiplication.  The fixed-point representation of the reciprocal has
1 sign bit and no integer bits.  Our \gls{fpga} implementation often
uses an on-\gls{fpga} \gls{dsp} block to execute this multiplication.
Twenty-five bits are used to represent the reciprocal as this is the
maximum width accepted by the \gls{dsp} modules on the \gls{fpga} used
for our error-rate simulations.  Theoretically, this results in a large
bit-width for the normalization output, however, unused bits are trimmed
during synthesis.

This normalization is trivial for certain variable degrees.  For
example, if $\card{\mathcal{N}_v(i)}$ is a power of 2, the
normalization can be implemented by bit-shifting the fixed-point
representation.  Similarly, if the reciprocal of
$\card{\mathcal{N}_v(i)}$ has few ones in its fixed-point
representation, soft logic can efficiently implement the resulting
multiplication. Thus, to simplify the normalization step, a
hardware-oriented code design approach can be taken where
$|\mathcal{N}_v(i)|$ is chosen to be a power of 2.

To form the \gls{vn} output $x_i$, the above normalization must be
projected onto the centered unit interval.  Similar to the
penalization step, the \gls{vn} tests whether or not the normalized
estimate is less than $-\frac{1}{2}$, greater than $\frac{1}{2}$, or
between $-\frac{1}{2}$ and $\frac{1}{2}$.  Two multiplexers are used
to set the variable estimate to be $-\frac{1}{2}$, $\frac{1}{2}$, or
the normalized estimate, respectively.

The final step of the \gls{vn} architecture is to format the variable
estimate $x_i$ to the correct fixed-point representation.  The
\gls{vtc} messages generally have a smaller bit-width than the
projected estimate.  Since the projected estimate is guaranteed to be
between $-\frac{1}{2}$ and $\frac{1}{2}$, its
fixed-point representation has 1 sign bit and no integer bits.
Therefore, only excess fraction bits need to be removed, which
causes the previously mentioned bit trimming for
the normalization output. While not indicated in Fig.~\ref{vnArch}, 
it is very important to round (rather than truncate) in order to remove
these fraction bits. Truncation (i.e., always rounding down) biases
decoding towards lower-weight codewords. Rounding prevents such a bias. 

From this description, one can observe that \admmlp \gls{vn}s are
simple to implement.  The most complex operation is the adder tree,
which gives \admmlp \gls{vn}s $\bigO{\card{\mathcal{N}_v(i)}}$ area
scaling and $\bigO{\log\card{\mathcal{N}_v(i)}}$ delay scaling.
Additionally, no information needs to be stored in the \gls{vn} for
use in future iterations. The result is a pipeline-friendly module.

%% file: checkNode.tex
\subsection{Check Node}

\begin{figure}[t]
	\centering
	\includegraphics[width=0.48\textwidth]{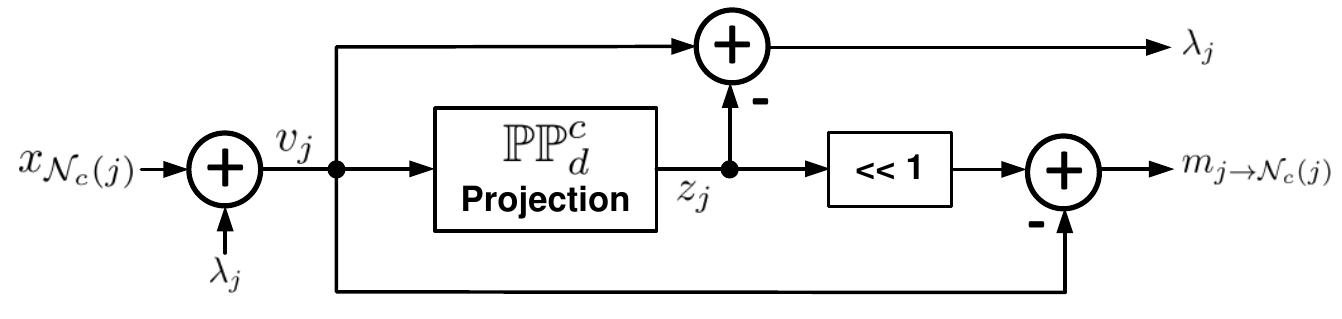}
	\FigCaption{Check node compute module.}
	\label{cnArch}
\end{figure}

Figure~\ref{cnArch} presents a schematic of a \gls{cn}, which executes
the operations on lines \ref{chkStart}--\ref{chkMsgUpLine} of Alg.~\ref{decodingAlg}.
A \gls{cn} first performs length-$\card{\mathcal{N}_c(j)}$ vector addition of the
incoming message vector $x_{\mathcal{N}_c(j)}$ with the check state
vector $\lambda_j$.  The \gls{vtc} messages and check states have the
same bit-width, but their fixed-point representations are different.
Therefore, to perform the vector addition, check states must be zero
extended to have the same number of fraction bits as the incoming
messages.  The length-$\card{\mathcal{N}_c(j)}$ vector addition output
$v_j$ has components with the same fixed-point representation as the
extended check states, except an additional integer bit is added to
prevent overflow.

The vector addition output is fed into the parity polytope projection
module.  It must also be temporarily stored while the projection takes
place so it can be used to calculate \gls{cn} outputs.  Implementation
of the parity polytope projection, the most resource intensive part of the \gls{cn},
will be covered in the next subsection.  
The replica variable vector $z_j$ is assigned the output of the
projection module.  The replica variable vector has the same bit-width
as the projection input, but its fixed-point representation has 1 sign
bit and no integer bits since its components are guaranteed to be in
$[-\frac{1}{2}, \frac{1}{2}]$.

Following parity polytope projection, new check state values and
outgoing messages $m_{j \rightarrow \mathcal{N}_c(j)}$ are calculated
in parallel using vector addition operations.  Before the check state
update, extra fraction bits are added to the vector addition result
$v_j$.  For the outgoing message calculation, the extended $v_j$ is
used with one fewer fraction bit since the parity polytope projection
is multiplied by 2.  This multiplication is accomplished by
bit-shifting the fixed-point representation of $z_j$.

The final step to output $\lambda_j$ and $m_{j \rightarrow
  \mathcal{N}_c(j)}$ from the \gls{cn} is to format their fixed-point
representations.  To discard excess integer bits, values are saturated
at the maximum or minimum that their representations allow.  Rounding
is performed to discard excess fraction bits.  This avoids the
aforementioned truncation-induced codeword biases in error-rate
performance.

Excluding projection onto the parity polytope, \gls{cn}s are simple to
implement.  The vector addition operations have constant delay in
check degree and $\bigO{\card{\mathcal{N}_c(j)}}$ area scaling.
\gls{cn} complexity lies in projection onto the parity polytope.
Projection gives \gls{cn}s
$\bigO{(\log\card{\mathcal{N}_c(j)})^2}$ delay scaling and
$\bigO{\card{\mathcal{N}_c(j)}(\log\card{\mathcal{N}_c(j)})^2}$
area scaling.  Storage of the  $v_j$ while the parity polytope
projection takes place occupies
$\bigO{\card{\mathcal{N}_c(j)}(\log\card{\mathcal{N}_c(j)})^2}$
area resources with
$\bigO{(\log\card{\mathcal{N}_c(j)})^2}$ pipeline stages.

%% file: ppArch.tex
\begin{figure}[htpb]
	\centering
        \includegraphics[width=0.48\textwidth]{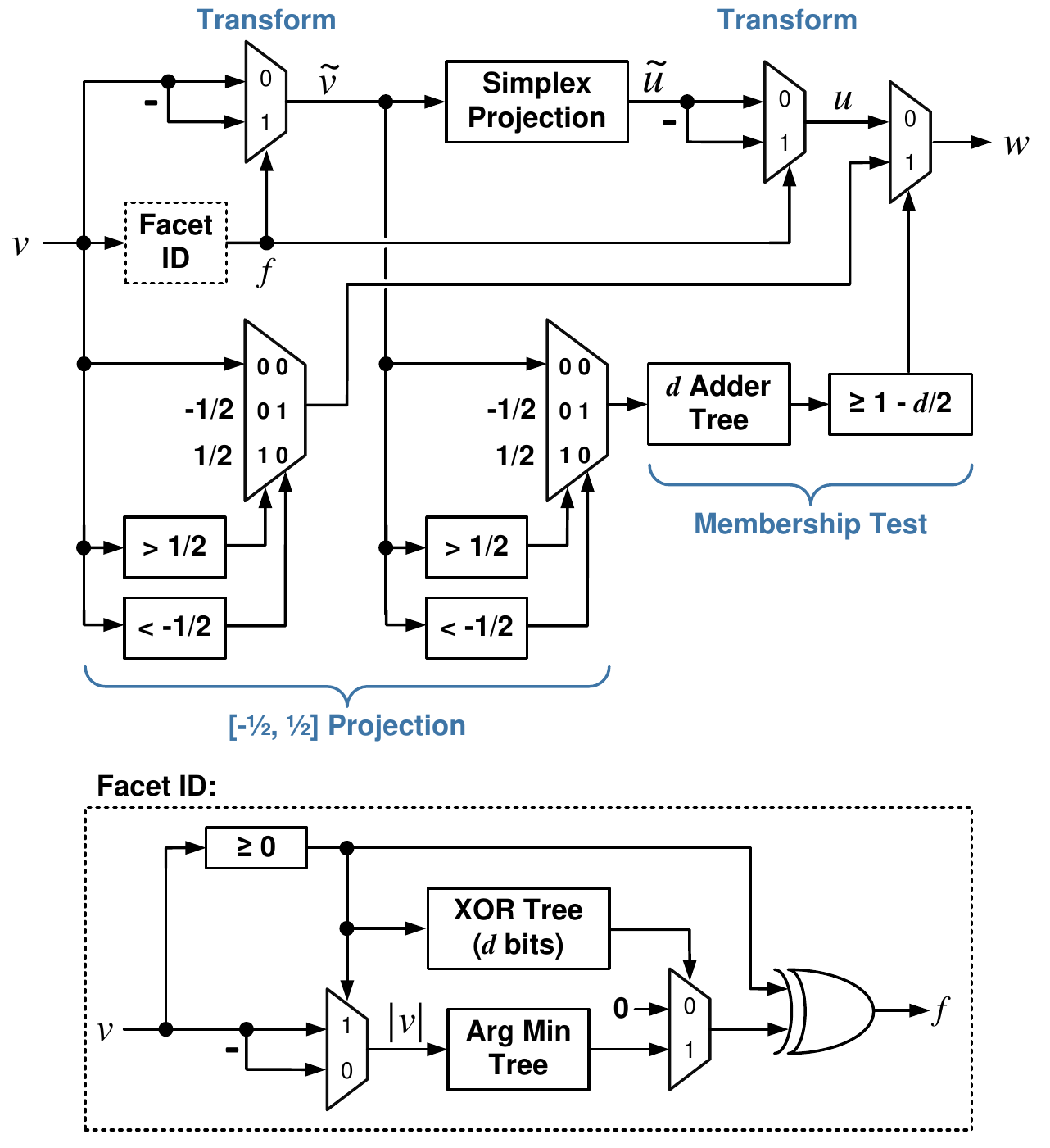}
        \FigCaption{Parity polytope projection. Note that the (dotted)
          facet identification module from the upper sub-figure is
          detailed in the lower sub-figure.}
	\label{ppArch}
\end{figure}

\subsection{Parity Polytope Projection}

Figure~\ref{ppArch} presents a schematic of the parity polytope
projection module.  As specified in Alg.~\ref{ppProjAlg}, the
operation starts with facet identification.  Finding the closest unit
hypercube vertex to the input vector $v$ is accomplished by checking
the sign bit of the fixed-point representation of $v$.  Finding the
closest odd-weight vertex is slightly more difficult.  First, a
length-$d$ vector corresponding to the absolute values of $v$ is
created with $d$ multiplexers choosing between $v_i$ and $-v_i$ for
each component of $v$.  This vector has the same fixed-point
representation as $v$ except the sign bit can be dropped since its
components are all non-negative.  This vector is fed into a min tree
that finds the minimum component and outputs a one-hot vector
indicating the index of the minimum.  If the closest vertex to $v$ is
even weight, the one-hot vector is used to flip the bit of $f$
corresponding to the minimum absolute value component via a
component-wise XOR.  The complexity of this operation lies in the min
tree which has $\bigO{\log d}$ delay and $\bigO{d}$ area scaling.
However, $v$ must be stored for $\bigO{\log d}$ pipeline stages,
resulting in $\bigO{d \log d}$ area usage.

With the active facet identified in $f$, a similarity transform is
executed on $v$ to align the active facet with the probability
simplex.  This is accomplished with $d$ multiplexers choosing between
$v_i$ or $-v_i$ based on the value of $f_i$.  The resulting vector
$\tilde{v}$ uses the same fixed-point representation as $v$, however,
since its components are guaranteed to be negative, the sign bit can
be dropped and added back in later when required for computation.
This operation has constant delay and linear area scaling in the
dimensionality of projection $d$.

At this algorithmic juncture there are three operations that can take
place in parallel: projection onto the unit hypercube, projection onto
the probability simplex, and testing parity polytope membership.
However, our implementation does not execute these operations in
parallel.  Parallel execution requires knowing the depth of each
operation in order to pad properly the lower latency operations with
pipeline registers.  This can not be done without knowing code check
degrees a priori.

Projection of $v$ onto the unit hypercube is the simplest operation to
perform.  For each component of $v$, two multiplexers choose between
$-\frac{1}{2}$, $\frac{1}{2}$, and $v_i$.  The fixed-point
representation is formatted to match the bit-width of the projection
output with 1 sign bit and zero integer bits.  This operation has
constant delay and linear area scaling in $d$.

Testing parity polytope membership involves projecting $\tilde{v}$
onto the unit hypercube.  Hypercube projection is performed in the
same manner as above.  This is followed by summing the resultant
vector using a minimum-depth adder tree.  In the adder tree, extra
integer bits are added to prevent overflow.  By comparing the adder
tree result to a constant, we are able to determine what the
projection output should be.  This decision is stored with a single
bit.  Due to the adder tree, this operation has $\bigO{d}$ area
scaling and $\bigO{\log d}$ delay scaling.

The implementation of simplex projection is the topic of the next
subsection.  Simplex projection dominates the complexity of parity
polytope projection.  It gives the parity polytope projection
$\bigO{d(\log d)^2}$ area scaling and $\bigO{(\log d)^2}$ delay
scaling.  Additionally, the hypercube projection of $v$ and the active
facet identifier $f$ must be stored for the $\bigO{(\log d)^2}$
pipeline stages it takes to execute the simplex projection.  This uses
$\bigO{d(\log d)^2}$ area.  The similarity transform is applied again
to the output of the simplex projection to invert itself.

The stored bit indicating parity polytope membership then drives $d$
multiplexers that choose to output the hypercube projection of $v$ or
the transformed output of the simplex projection module.  Both of
these possible outputs have fixed-point representations with 1 sign
bit and no integer bits.

%% file: simpArch.tex
\subsection{Simplex Projection}

\begin{figure}[t]
	\centering
	\includegraphics[width=0.48\textwidth]{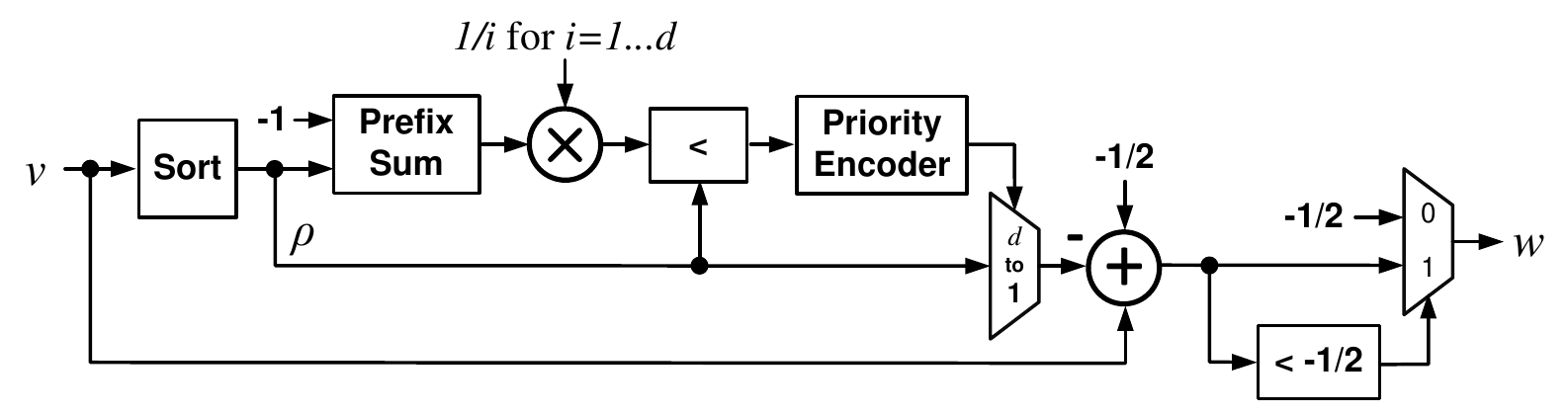}
	\FigCaption{Simplex projection.}
	\label{simpArch}
\end{figure}

Our algorithm for simplex projection is detailed in
Alg.~\ref{simpProjAlg}; a schematic is depicted in
Fig.~\ref{simpArch}. The components of the vector to be projected are
first sorted in descending order.  To sort in hardware, we require the
set of operations executed to be performed regardless of the input
vector.  Sorting networks accomplish this.  Sorting networks are
composed of compare-swap modules, each of which can be implemented
with a compare operation and two multiplexers.  We implement
delay-optimal sorting networks from Knuth~\cite{knuth_1998_tacp}.

%Knuth provides an in-depth analysis of sorting
%networks~\cite{knuth_1998_tacp}.  He explains that the best
%\tcb{practical} \tcr{(``practical''?)}  sorting networks achieve
%$\bigO{(\log d)^2}$ delay scaling and $\bigO{d(\log d)^2}$ area
%scaling.  This optimal scaling is obtained by Batcher's odd-even merge
%sort~\cite{batcher_1968_merge} which we implement
%in~\cite{wasson_2015_pp,wasson_2016_thesis}.  Batcher's sort uses a
%special merge method that recursively merges the odd and even indexed
%elements of two separately sorted vectors.  The result is two new
%vectors that can be merged with one layer of compare and swap modules.
%This yields $\bigO{(\log d)^2}$ delay scaling since each merge
%requires $\bigO{\log n}$ layers, and $\bigO{\log n}$ merges are
%required.  The area scaling of $\bigO{d(\log d)^2}$ is achieved since
%$\bigO{d}$ compare-swap modules are required in each layer.  We note
%that the input vector $v$ must be stored, requiring $\bigO{d(\log
%  d)^2}$ area.  Knuth~\cite{knuth_1998_tacp} provides delay optimal
%sorting networks for $d \leq 16$.  We use these optimal networks in
%our implementation.

The next step of simplex projection is to calculate all partial sums
(termed ``prefix sum'') of the sorted vector $\rho$.  Since we need to
subtract 1 from every partial sum, we simply include $-1$ as part of
the prefix sum input.  The prefix sum operation can be performed with
$\bigO{d}$ area scaling and $\bigO{\log d}$ delay scaling.  Ladner and
Fischer describe such a construction~\cite{ladner_1980_prefix}.  The
$d^{th}$ sum is computed with a minimum-depth adder tree.  Other sums
are calculated by reusing computations when possible, making linear
area scaling possible.  Extra integer bits are allocated to the
fixed-point representation of the prefix sum output to prevent
overflow.  Note that both $v$ and $\rho$ must be stored during this
operation, requiring $\bigO{d\log d}$ area.

Next, the prefix sum output vector components are normalized by their
respective component indices.  Component index reciprocals are found
during synthesis and the normalization is performed by multiplication.
As with \gls{vn}s, multiplication by a power of 2 can easily be
implemented with FPGA soft logic for some indices, while a multiplier
DSP core is required for others.  This operation has constant delay
and linear area scaling in the dimension of projection.

We wish to select the normalized partial sum with the largest index
that satisfies $\rho_i>u_i$ as the common shift in the simplex
projection.  First, a length-$d$ binary vector is created indicating
the indices satisfying $\rho_i>u_i$.  A priority encoder is then used
to create a one-hot vector indicating the largest index position
satisfying $\rho_i>u_i$.  This is also a prefix
operation, which yields the same complexity as the prefix
sum~\cite{wasson_2015_pp}.  However, the operation is on a binary
vector, and not a fixed-point vector.  The resources consumed are thus
much smaller.  This one-hot vector is used to select the corresponding
component of $u$.

Finally, the selected component $u_{i^{\ast}}$ and $\frac{1}{2}$ are
subtracted from all components of $v$.  This is accomplished with two
adders for every component of $v$.  After this, each component is
compared to $-\frac{1}{2}$ and a multiplexer chooses between the adder
results and $-\frac{1}{2}$ to form the final output.  The output
bit-width matches the input bit-width.  The output fixed-point
representation has 1 sign bit and no integer bits since each component
is between $-\frac{1}{2}$ and $\frac{1}{2}$.

%% file: results.tex
%\section{Results}
In order to test the hardware viability of \admmlp decoding, we use an
\gls{fpga}-in-the-loop simulation environment that consists of a PCI-based 
Xilinx Virtex-5 \gls{fpga} platform on \gls{pc}.  The proposed
architecture was synthesized on the \gls{fpga}, along with the wrapper
logic needed for noise generation and data transfer to a software test
bench.

The binary-input \gls{awgn} channel is simulated using a Gaussian
random number generator~\cite{liu_2015_rng} on the \gls{fpga}. The
core is a linear feedback shift register of period $2^{176}$, fed into
an approximation of the inverse cumulative distribution function.
Channel simulation was performed on the \gls{fpga} to minimize
simulation time by eliminating the bottleneck of PC-to-FPGA data
transfer.  We verified that FPGA-based channel simulation produced the
same \gls{fer} results as CPU-based channel simulation for
low-\gls{snr} channels.

Three \qcldpc codes are considered for error-rate simulation and
resource usage analysis. The first is the [155, 64, 20] Tanner
code~\cite{tanner_2001_refcode}, whose parity-check matrix is composed
of $31\times 31$ cyclic matrices.  The second is the [672, 546] WiGig
code~\cite{IEEE-80211ad_rev} composed of $42\times 42$ matrices.  The
final codes are an ensemble of five [1002, 503] (3,6)-regular \qcldpc
codes.  The five parity-check matrices for this ensemble were created
by randomly generating shifts for $167 \times 167$ identity matrices.
The resulting factor graph girths were verified using techniques
from~\cite{draper_2013_qc}.  Codes with girth less than 6 were
discarded.  Example shift matrices are provided in
Fig.~\ref{matrixFig}.

\begin{figure}[t]
	\begin{equation*}
	\left[\begin{smallmatrix}
	30 & 29& 27& 23& 15 \\
	26 & 21& 11&  22&  13\\
	6 & 12& 24& 17& 3 \\
	\end{smallmatrix}\right]
	%  Its
	%parity check matrix consists of $42\times 42$ shifted identity
	%matrices and all-zeros matrices.  The matrix shifts are given by
	%where ``$-$'' denotes an all-zeros matrix.   
	%The parity-check
	%matrices are composed of $167 \times 167$ shifted identity matrices.
	%Parity-check matrices for this ensemble were created by randomly
	%generating the identity matrix shifts.  Then, using techniques
	%from~\cite{draper_2013_qc}, the resulting factor graph girths were
	%checked.  Codes with girth less than 6 were discarded.  An example
	%base matrix from this ensemble is
	\left[\begin{smallmatrix}
	115 & 13& 25& 166& 17&129 \\
	124 & 38& 137& 13&  160 & 136\\
	75 & 152& 89& 73& 0 & 145 \\
	\end{smallmatrix}\right] 
	\end{equation*}
	\begin{equation*}
	\left[\begin{smallmatrix}
		29& 30& 0& 8& 33 & 22&17&4&27&28&20&27&24&23&-&- \\
		37& 31& 18&  23&  11&21&6&20&32&9&12&29&10&0&13&-\\
		25& 22& 4& 34& 31&3&14&15&4&2&14&18& 13&13&22&24\\
	\end{smallmatrix}\right]
	\end{equation*}
	
	\caption{Shifts for \gls{qc} parity check matrices where
	``$-$'' denotes an all-zeros matrix. Clockwise from top left:
	Tanner code, (3, 6) ensemble, WiGig code.}  \label{matrixFig}
\end{figure}

Before decoders for these codes are implemented, design decisions
regarding the fixed-point representation of messages must be made.
The first decision is the input/output bit-width of the decoder.  An
input/output bit-with of 8 bits was required to guarantee error-rate
performance close to double-precision
implementations~\cite{wasson_2016_thesis}.  Fewer bits can be used at
the cost of deteriorating error-rate performance. However, the rate of
deterioration depends on the code.  For example, we found the WiGig
code \gls{fer} performance to be extremely sensitive to decreasing
bit-width.  Next, the number of bits used for internal messages needs
to be decided.  Recall that the main effect of these bits is to
provide \gls{ctv} messages the additional dynamic range needed to
override channel information.  It was found that 2 additional bits are
required to provide good performance in higher-reliability
channels~\cite{wasson_2016_thesis}.  Next, the bit allocations for
\gls{llr}s, \gls{ctv} messages, and check states must be determined.
In our architecture, these three values all use the same number of
fraction bits.  Experimentation indicates that maximizing the number
of fraction bits results in the best error-rate
performance~\cite{wasson_2016_thesis}.  That is, \gls{llr}s should
have no integer bits, and the \gls{ctv} messages and check states
should have 2 integer bits. We use these allocations in all our
simulations.

Table~\ref{table:Fxpt-Messages} summarizes the fixed-point precision
of each message variable as determined through experimentation to
obtain FER performance close to double-precision. Message variables
are grouped by computation module and expressed as signed fixed-point
numbers in the Q format \cite{khan2011digital}.

\begin{table}[t]
\begin{center}
\caption{Fixed-Point Message Characteristics (msg / bit-width / format)}
\label{table:Fxpt-Messages}
\scalebox{1}{
\renewcommand{\arraystretch}{1}
\begin{tabular}{!{\vrule width 3\arrayrulewidth} C{0.8in} | C{0.2in} | C{0.3in} !{\vrule width 3\arrayrulewidth} C{0.5in} | C{0.2in} | C{0.3in} !{\vrule width 3\arrayrulewidth}}
\Xhline{3\arrayrulewidth} 

\multicolumn{3}{!{\vrule width 3\arrayrulewidth} c !{\vrule width 3\arrayrulewidth}}{\textbf{Variable Node}} & \multicolumn{3}{ c !{\vrule width 3\arrayrulewidth} }{\textbf{Simplex Projection}} \\ \hline
\Xhline{3\arrayrulewidth}
		$\gamma_i$ & 8 & Q0.7 & $v$ & 14 & Q4.9 \\ \hline
		$t_i$ & 12 & Q4.7 & $\rho$ & 14 & Q4.9 \\ \hline
		$s_i$ & 13 & Q5.7 & $w$ & 14 & Q0.13 \\ \hline
\Xhline{3\arrayrulewidth}

\multicolumn{3}{!{\vrule width 3\arrayrulewidth} c !{\vrule width 3\arrayrulewidth}}{\textbf{Check Node}} & \multicolumn{3}{ c !{\vrule width 3\arrayrulewidth}}{\textbf{Parity Polytope Projection}} \\ \hline
\Xhline{3\arrayrulewidth}
		$x_i$, $x_{\mathcal{N}_c(j)}$ & 10 & Q0.9 & $v$ & 13 & Q3.9 \\ \hline
		$m_{\mathcal{N}_v(i)\rightarrow i}$, $m_{j \rightarrow \mathcal{N}_c(j)}$, $\lambda_j$ & 10 & Q2.7 & $\tilde{v}$ & 14 & Q4.9 \\ \hline
		$v_j$ & 13 & Q3.9 & $\tilde{u}$ & 14 & Q0.13 \\ \hline
		$z_j$ & 13 & Q0.12 & $w$ & 13 & Q0.12 \\ \hline
\Xhline{3\arrayrulewidth}

\end{tabular}
}
\end{center}
\end{table}

An additional parameter that affects error-rates and resource
utilization is the number of decoding iterations.  Similar to
\gls{bp}, \admmlp can be configured to terminate after a maximum
number of iterations.  This can enforce latency and throughput
constraints.  Experimentation found that at least
60 iterations are required for our fixed-point configuration to achieve
error-rate performance close to its capabilities without a limit on the
maximum number of iterations~\cite{wasson_2016_thesis}.

%The Verilog source code implementing the proposed decoder is available
%online~\cite{WassonCode}.

%% file: errorResults.tex
\subsection{Error-Rate Performance}
The previously mentioned parameter choices affect both error-rate
performance and resource consumption.  There are two additional
parameters that only affect error-rate performance.  We discuss the
choice of these parameters here.

Simulated channel outputs need to be saturated at some value in order
to produce LLRs within the decoder's input range.  We parameterize
this in terms of standard deviations of channel noise.  That is, the
channel output is saturated at $\pm(1+a\sigma)$ where $a>0$ and
$\sigma$ is the standard deviation of the added Gaussian noise.
Experimentation revealed our implementation is not extremely sensitive
to this parameter, but $a=1$ was found to be optimal with respect to
\gls{fer}~\cite{wasson_2016_thesis}.  Therefore, we saturate channel
outputs one standard deviation beyond the transmission values $\pm 1$.
The saturated channel outputs are then scaled such that the saturation
values are mapped to minimum and maximum \gls{llr} values.  Recall
that \gls{awgn} channel outputs are proportional to \gls{llr} values,
and scaling \gls{llr}s does not change the \gls{lp} decoding
objective.

The final parameter configuration is to choose a suitable penalty
parameter $\alpha$.  The optimal penalty parameter changes with
respect to SNR where larger penalty parameters perform better on
low-\gls{snr} channels, while smaller penalty parameters perform better
on high-\gls{snr} channels.  A penalty parameter of $\alpha=0.1$ was
found to give good performance across the tested channels for all
three codes~\cite{wasson_2016_thesis}.  

\begin{figure*}[t]
	\centering 
		\subfloat[Tanner code]{\includegraphics[width=.32\textwidth]{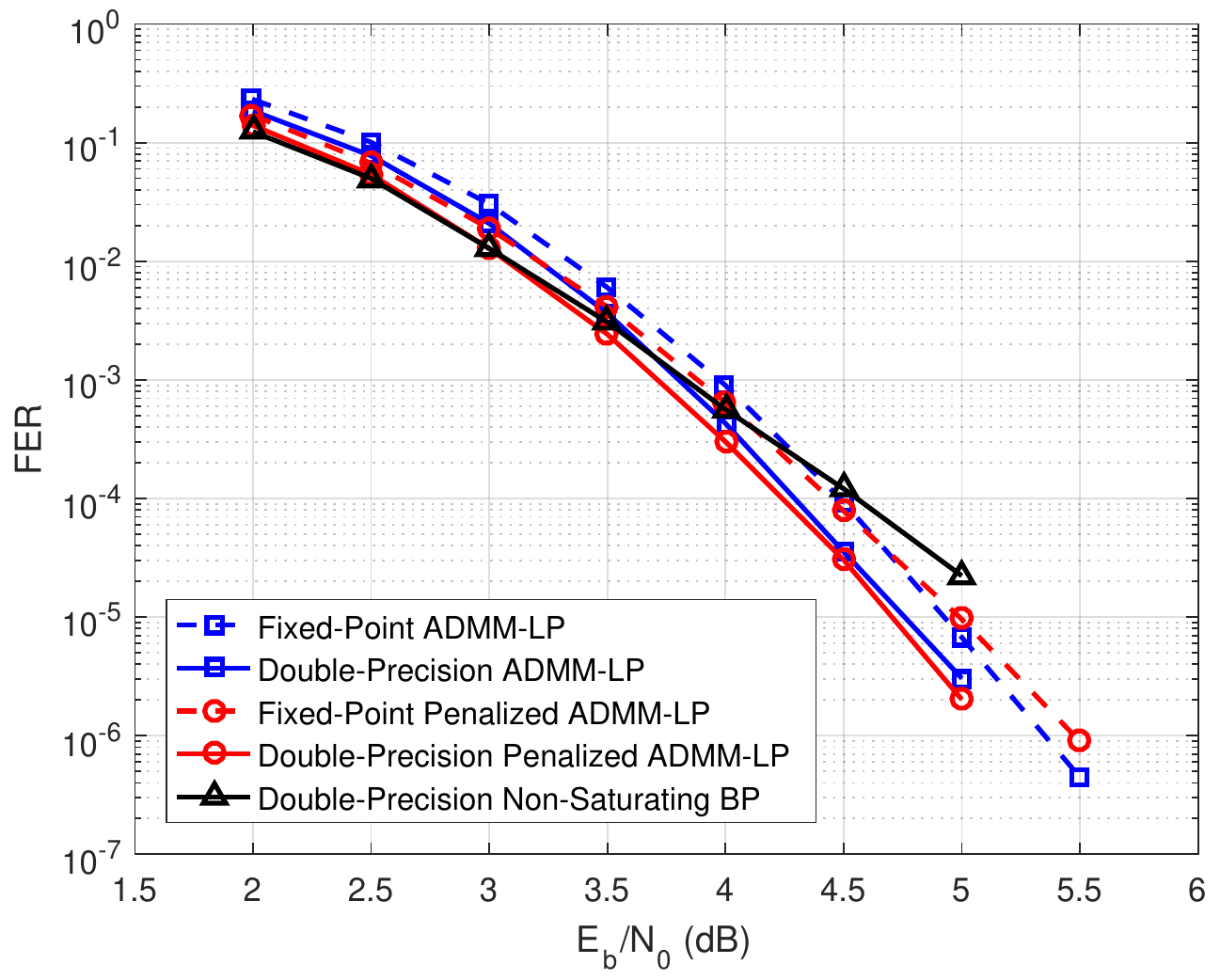}\label{tannerFER}}
        \hfill 
        \subfloat[WiGig code]{\includegraphics[width=.32\textwidth]{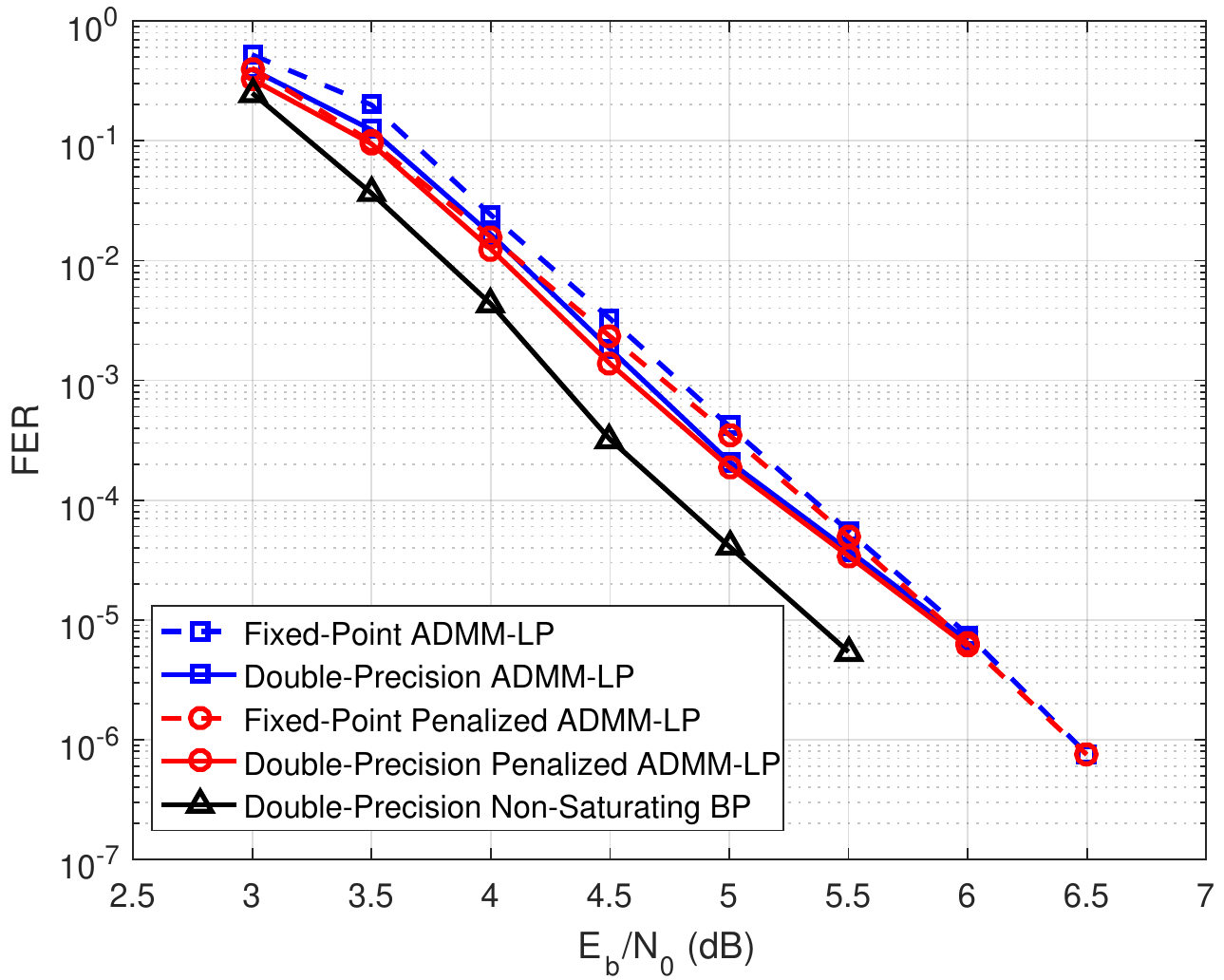}\label{wigigFER}}
        \hfill 
        \subfloat[\qcldpc ensemble]{\includegraphics[width=.32\textwidth]{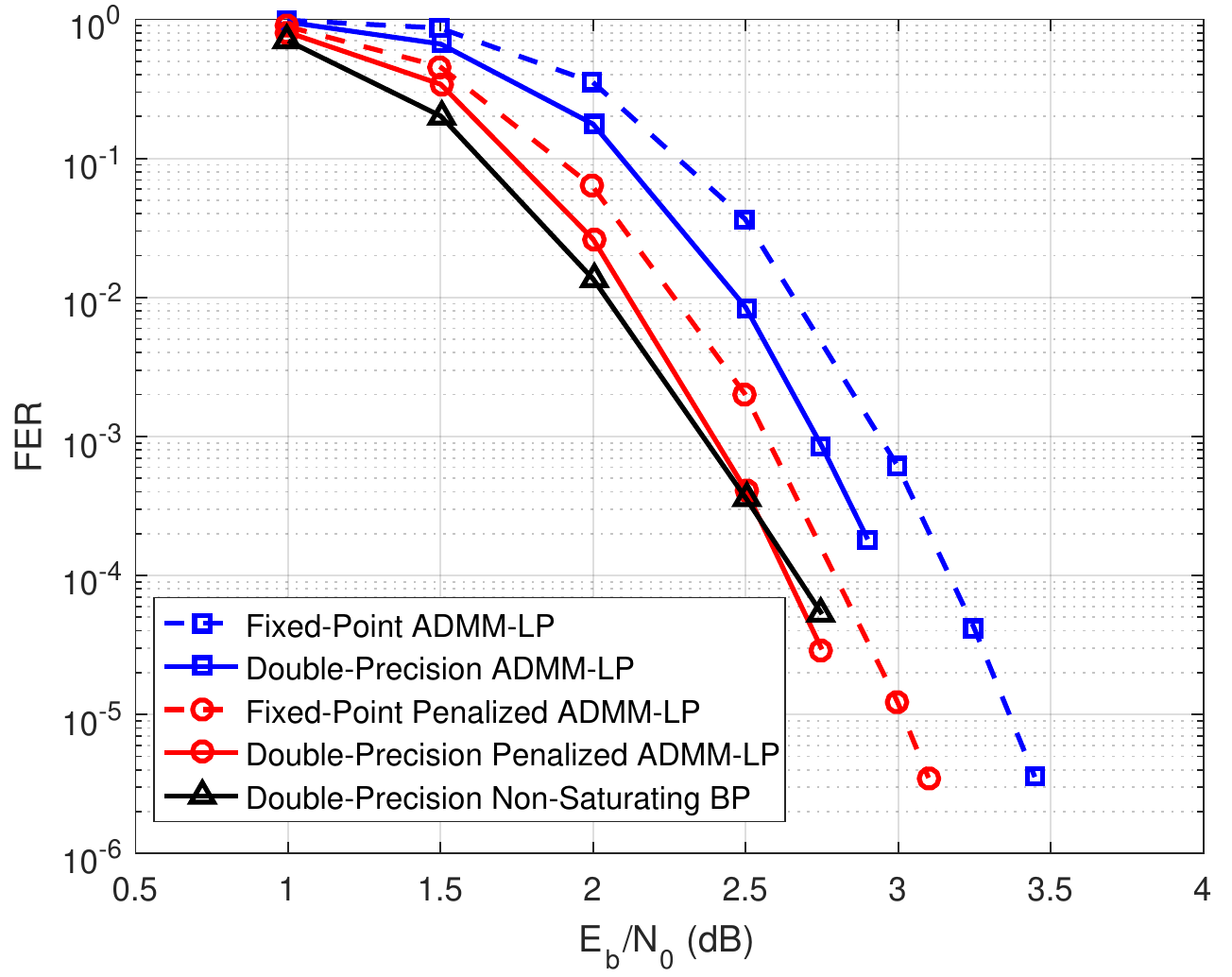}\label{ensembleFER}}
        \hfill 
        \FigCaption{\gls{fer} performance of the Tanner, WiGig, and \qcldpc ensemble codes.}
	\label{FERresults}
\end{figure*}

Figure~\ref{FERresults} presents the \gls{fer} experimental results
for the three codes under investigation on the binary input \gls{awgn}
channel.  We present results for both penalized ($\alpha=0.1$) and
unpenalized ($\alpha=0$) \admmlp, where the value of $\alpha$ refers
to its setting in Alg.~\ref{LPDecoding} after the reparameterization
that eliminated $\mu$.  Double-precision Sum-Product \gls{bp} and
\admmlp results are also plotted to form a basis for comparison.  Each
point on the following plots represents an accumulation of 100 frame
errors.  Double-precision simulations for \admmlp and \gls{bp} were
performed using Liu's implementation~\cite{liu_2013_dp}.  The \gls{bp}
results shown were generated with Butler and Siegel's non-saturating
version described in~\cite{siegel_2014_bp}.  The same limit of 60
iterations is used for all decoding algorithms.

%Each of the FER performance figures presented in this section displays
%the fixed-point and double-precision performance for both the
%penalized and unpenalized \admmlp performance alongside
%double-precision \gls{bp}.

%\begin{figure}[t]
%	\centering
%	\includegraphics[width=0.48\textwidth]{./figures/MM_FER_curves/FER_Combined_Tanner.pdf}
%	\FigCaption{\gls{fer} performance of the Tanner code.}
%	\label{tannerFER}
%\end{figure}
%
%\begin{figure}[t]
%	\centering
%	\includegraphics[width=0.48\textwidth]{./figures/MM_FER_curves/FER_Combined_WiGig.pdf}
%	\FigCaption{\gls{fer} performance of the WiGig code.}
%	\label{wigigFER}
%\end{figure}
%
%\begin{figure}[t]
%	\centering
%	\includegraphics[width=0.48\textwidth]{./figures/MM_FER_curves/FER_Combined_Ensemble.pdf}
%	\FigCaption{\gls{fer} performance of the \qcldpc ensemble.}
%	\label{ensembleFER}
%\end{figure}

\subsubsection{Tanner Code}

Fig.~\ref{tannerFER} presents the FER performance of the Tanner code.
A small performance gap exists between the fixed-point and
double-precision \admmlp implementations. At higher \gls{snr}s, all
\admmlp implementations outperform double-precision \gls{bp}. The
penalized \admmlp decoder closes the gap to double-precision \gls{bp}.
However, it does not perform as well as unpenalized \admmlp at
$E_b/N_0=5.5$dB. These results support the conclusion that unpenalized
\admmlp is better suited to high-\gls{snr} channels.

\subsubsection{WiGig Code}

Fig.~\ref{wigigFER} displays the FER performance of the WiGig
code. Again, we see that in both cases, fixed-point \admmlp maintains
very close performance to the double-precision implementation with
10-bit messages. However, there is a very large performance gap
between \gls{bp} and ADMM-LP. This performance gap is not closed with
the addition of penalization; the opposite of what has been observed
with other codes~\cite{draper_2014_penalized_lp}. The root cause of
the weakness of LP decoding for this code requires further
investigation.  We conjecture that, in part, it may be due to the
high-degrees of the check nodes, resulting in more pseudocodewords, or
due to the variable-one variable nodes.

%\tcr{Stark: was this sentence you? if so, no problem. if not, then
%  remove it because i have no idea whether or not this would make an
%  impact.}

\subsubsection{\qcldpc Ensemble}
The FER performance of the ensemble of $(3,6)$-regular \qcldpc codes is shown in
Fig.~\ref{ensembleFER}.  Each curve is obtained by averaging the
performance of the same five codes from the \qcldpc ensemble.  This
experiment is a more powerful demonstration of the performance of
ADMM-LP, where the addition of penalization closes the large
performance gap between \gls{bp} and ADMM-LP.  The fixed-point
implementations of both penalized and unpenalized \admmlp achieve
performance very close to double-precision.

%% file: resourceResults.tex
\subsection{Resource Usage}

This section examines the FPGA resource utilization for the three
decoders synthesized for the Tanner, WiGig, and QC-LDPC ensemble codes
using the fixed-point message representations summarized in
Table~\ref{table:Fxpt-Messages}.  While the error-rate simulations
were performed on an older Xilinx Virtex-5 \gls{fpga}, the resource
utilization results presented here target a newer Altera Stratix V
\gls{fpga} (model 5SGXEA7N2F45C2). This \gls{fpga} has 234,720
\gls{alm}, 256 \gls{dsp} blocks, and 2,560 ``M20K'' \gls{ram} blocks
with 52,428,800 \gls{ram} bits in total. Synthesis was performed using
Altera's Quartus II (15.0.0) tool suite using balanced
optimization. Basic power estimation was performed using gate-level
simulations of high-noise decodings and Altera's power analyzer tool.
Table~\ref{table:Code-Resource-Comparison} summarizes the FPGA
utilization and throughput results.

\begin{table}[ht]
\begin{center}
\caption{FPGA performance results and resource comparison}
% on Altera Stratix V for three codes under investigation}
\label{table:Code-Resource-Comparison}
\scalebox{0.98}{
\renewcommand{\arraystretch}{1}
\begin{scriptsize}
\begin{tabular}{| C{1.2in} | C{0.53in} | C{0.53in} | C{0.53in} |}
\Xhline{3\arrayrulewidth}
\textbf{Specification} & \textbf{Tanner}   & \textbf{WiGig}  & \textbf{Ensemble}  \\ \hline                     
\Xhline{3\arrayrulewidth}

\textbf{Block Length} & 155 & 672 & 1002 \\ \hline
\textbf{Code Rate} & 2/5 & 13/16 & 1/2 \\ \hline
\textbf{QC Matrix~Structure ($r$~rows,~$s$~columns)} & $r = 3$ $s=5$ & $r = 3$ $s=16$ & $r = 3$ $s=6$ \\ \hline
\textbf{Expansion Factor} & 31 & 42 & 167 \\ \hline
\Xhline{3\arrayrulewidth}

\multicolumn{4}{|c|}{\textbf{FPGA Performance Results}}\\ \hline
\Xhline{3\arrayrulewidth}
\textbf{Clock Freq. (MHz)} & 237 & 221 & 225 \\ \hline
\textbf{Iterations} & 60 & 60 & 60 \\ \hline
\textbf{Cycles Per Iter.} & 137 & 189 & 440 \\ \hline
\textbf{Throughput (Mb/s)} & 4.47 & 13.16 & 8.52 \\ \hline
\textbf{Latency Per Iter. ($\mu$s)} & 0.578 & 0.851 & 1.960 \\ \hline
\textbf{Total Est. Power (mW)} & 758 & 2153 & 863 \\ \hline
\textbf{Dynamic Power (mW)} & 703 & 1942 & 797 \\ \hline
\textbf{Static Power (mW)} & 55 & 211 & 66 \\ \hline
\textbf{Adaptive Logic Modules} & 12201 & 34715 & 14315 \\ \hline
\textbf{DSP Blocks} & 11 & 46 & 15 \\ \hline
\textbf{Memory (Bits)} & 16,430 (40~BRAMs) & 67,284 (92~BRAMs) & 106,212 (47~BRAMs) \\ \hline
\Xhline{3\arrayrulewidth}

\multicolumn{4}{|c|}{\textbf{FPGA ALM Resource Utilization}}\\ \hline
\Xhline{3\arrayrulewidth}
\textbf{Check Nodes} & 85\% & 94\% & 85\% \\ \hline
\textbf{Variable Nodes} & 5\% & 3\% & 6\% \\ \hline
\textbf{Memory} & 8\% & 2\% & 8\% \\ \hline
\textbf{Control Logic} & 2\% & 1\% & 1\% \\ \hline
\Xhline{3\arrayrulewidth}

\multicolumn{4}{|c|}{\textbf{Maximum Number of Pipeline Stages}}\\ \hline
\Xhline{3\arrayrulewidth}
\textbf{Check Node} & 46 & 54 & 47 \\ \hline
\textbf{Variable Node} & 9 & 9 & 9 \\ \hline
\Xhline{3\arrayrulewidth}

\multicolumn{4}{|c|}{\textbf{FPGA Power Breakdown}}\\ \hline
\Xhline{3\arrayrulewidth}
\textbf{Check Nodes} & 76\% & 84\% & 76\% \\ \hline
\textbf{Variable Nodes} & 7\% & 4\% & 6\% \\ \hline
\textbf{Memory} & 17\% & 12\% & 17\% \\ \hline
\Xhline{3\arrayrulewidth}

\end{tabular}
\end{scriptsize}
}
\end{center}
\end{table}

\subsubsection{Tanner Code}

The implementation of the Tanner code decoder in the partially-parallel
architecture has three degree-5 CNs and five degree-3 VNs. Each
\gls{vn} uses a single \gls{dsp} block to accomplish its
normalization, and each \gls{cn} uses two \gls{dsp} blocks to perform
division by 3 and division by 5.

From Table~\ref{table:Code-Resource-Comparison}, we see that CNs
account for the majority of resource usage.  Therefore, a further
breakdown of CN resource consumption is warranted. We now break down
the \gls{alm} and power consumption inside a \gls{cn} on a
sub-component basis.  Parity polytope projection accounts for 89\% of
\gls{alm} and 91\% of power usage inside the check node. Simplex
projection accounts for a bit over 51\% and 54\%, respectively.
Finally, sorting consumes 14\% of \gls{alm} and power usage, and
prefix addition consumes 11\% of \gls{alm} and power usage.  Note that
these figures are nested. For example, the 91\% of \gls{cn} power
usage attributed to parity polytope projection includes power used in
simplex projection.  We believe this is due to heavy \gls{alm} usage
for intermediate storage.  For example, inside a \gls{cn}, $v_j$ must
be stored until projection is complete.  In our implementation, the
resources used for this storage count toward polytope projection
resource consumption.  Since area utilization and power are related,
it is not surprising that the ALM and power breakdowns are quite
similar.

\subsubsection{WiGig Code}
The WiGig code has one degree-16 \gls{cn}, one degree-15 \gls{cn}, and
one degree-14 \gls{cn}.  It has 14 degree-3 \gls{vn}s, one degree-2
\gls{vn}, and one degree-1 \gls{vn}. The degree-3 VNs use fewer
resources than the Tanner decoder due to increased resource sharing
among the 14 degree-3 VNs.

Again, \gls{cn}s account for the majority of resource usage.  The
percentage of \gls{alm} usage and power consumption inside the
degree-16 \gls{cn} on a sub-component basis are very similar to the
degree-5 \gls{cn}s of the Tanner decoder.  For the WiGig code, the two
complexity-dominating operations, sort and prefix addition, consume a
larger fraction of resources.

\subsubsection{\qcldpc Ensemble}
The QC-LDPC ensemble implementation has six degree-3 \gls{vn}s and
three degree-6 \gls{cn}s.  CNs account for the majority of resource
usage, and the internal \gls{cn} resource breakdown is again almost
identical to that of the Tanner code \gls{cn}s.  The same trend has
emerged for all decoders, where pipeline depth and intermediate value
storage have a large impact on resource consumption.

%% file: implementationComp.tex
\begin{table}%[ht]
%\begin{center}
\caption{Comparison of FPGA-based LDPC decoder implementations}
\label{table:FPGA-decoder-comparison}
\scalebox{1}{
\renewcommand{\arraystretch}{1}
\hspace*{-0.8em}\begin{tabular}{!{\vrule width 3\arrayrulewidth} C{0.52in} !{\vrule width 3\arrayrulewidth} C{0.28in} | C{0.28in} | C{0.45in} | C{0.43in} | C{0.43in} !{\vrule width 3\arrayrulewidth}}
\Xhline{3\arrayrulewidth}
& \textbf{\cite{Lei2006_comp}}                   & \textbf{\cite{Zarubica2007_comp}}                     & \textbf{\cite{Bhagawat2005_comp}} & \textbf{\cite{Chandrasetty2011_comp}}                     & \textbf{This} \\                      
& 2006                     & 2007                     & 2005 & 2011                     & \textbf{Paper}                                                     \\
\Xhline{3\arrayrulewidth}
\textbf{Alg}                                                                               & \gls{bp}                 & \scriptsize{Min-Sum}                 & N/A & Min-Sum                  & \scriptsize{ADMM-LP}                                               \\ \hline
\textbf{Arch}                                                                                      & Serial                     & FP               & PP                           & PP                    & PP                        \\ \hline
\textbf{Length}                                                                                & 980                    & 1200                & 1038  & 1152                     & 1002 \\ \hline
\textbf{Dsn Rate} & 0.696    & 1/2              & 1/2       & 1/2        & 1/2 \\ \hline
\textbf{Struct}                                                                                                          & N/A                      & PEG     & QC   & QC & QC                                                       \\ \hline
\textbf{Max Iter}  & 100   & 10   & 18 & 10    & 60    \\ \hline
\textbf{Perf @ $\text{3dB}$ } & $\text{BER}$  \hspace*{-0.7em} $8   \tiny{\times} \! 10^{-5}$ & $\text{FER}$  \hspace*{-0.3em}$5 \tiny{\times} \! 10^{-2}$ & N/A & $\text{BER}$ $1 \tiny{\times} \! 10^{-5}$ & \scriptsize{$\text{FER}$ $1 \tiny{\times} \! 10^{-5}$ $\text{BER}$ $2 \tiny{\times} \! 10^{-7}$} \\ \hline
\Xhline{3\arrayrulewidth}
\textbf{Device}   & AC-EP1C6 &	XV-4 &	XV-E &	XV-2	 & AS-V \\ \hline
\textbf{Msg Bit Width }                                                                                          & 6                       & 3              & 4  & 4                   & LLR 8 Int. 10                                                              \\ \hline
\textbf{\scriptsize{Early Term}}       & No      & No      & No  & No       & No \\ \hline
\textbf{\scriptsize{Freq (MHz)}} & 136 & 100 & 26 & 64 & 224 \\ \hline
\textbf{\scriptsize{Thpt (Mb/s)}} &  7	& 6000	& 72 & 50	& 8.52 \\ \hline
\textbf{Thpt/Iter (Mb/s)} & 0.07 &	600	& 4 & 5	& 0.142 \\ \hline
\textbf{Delay / Iter ($\mu$s)} & 1.40	& 0.02	& 0.80 & 2.30	& 1.96 \\ \hline
\textbf{\scriptsize{Pwr (mW)}} & N/A	& N/A	& 322	 & N/A & 863 \\ \hline
\textbf{Resources} & 997 ALMs	& 40613 Slices	& 10883 Slices & 2778 Slices & 14315 ALMs \\ \hline
\textbf{Mem (Kbits)} & 34  & N/A &N/A: \hspace*{-0.5em}120~BRAMs& 19.5Kb: \hspace*{-0.25em}29~BRAMs & 106.2Kb: \hspace*{-0.25em}47~BRAMs \\ \hline 
\Xhline{3\arrayrulewidth}
\end{tabular}
}
%\end{center}

% Footnote
\vspace{0.75ex}
\scalebox{1}{
\hspace{-1em}\begin{tabular}{L{8.75cm}}
Note that the BER presented here corresponds to that achieved by
fixed point penalized ADMM-LP at $\text{FER}=1.2 \! \times \!
10^{-5}$ as plotted in Fig.~\ref{ensembleFER}.  The acronyms used are as follows:
FP = ``fully parallel'', PP = ``partially parallel'', PEG =
``progressive edge growth'', AE = ``Altera Cyclone'', \ \ XV = ``Xilinx
Virtex'', AS = ``Altera Stratix''.
\end{tabular}
}

\end{table}

\subsection{Implementation Comparison}

Table~\ref{table:FPGA-decoder-comparison} compares our implementation
for the \qcldpc ensemble with several FPGA-based \gls{ldpc} decoders
having similar code rates and comparable block length.

Our ADMM-LP decoder achieves better error-correction performance at
the chosen $E_b/N_0=3$dB operating point compared to the four
comparison works as \admmlp outperforms Min-Sum in this metric. On the
other hand, our decoder requires more iterations and larger bit
widths, resulting in in lower throughput and higher logic utilization.

%Decoder throughput and power consumption could be improved by
%implementing an early-termination scheme with additional logic in
%each CN module to reduce the number of decoding iterations without
%sacrificing error-correction performance.

A direct comparison of logic resources and overall area between
designs implemented on Altera and Xilinx FGPAs is nearly impossible.
The internal lookup table and D-flipflop structure of a Xilinx Slice
is not equivalent to an Altera ALM~\cite{alteraALMxilinxSlice}, and
the ALM and Slice architectures change from one FPGA generation to
another.  Additionally, FPGA synthesis and place-and-route stages are
highly dependent on the target device. Nevertheless, the logic
utilization numbers of Table~\ref{table:FPGA-decoder-comparison} show
that our partially-parallel ADMM-LP decoder implementation has a logic
resource utilization within an order of magnitude of the
partially-parallel comparison works implemented on Xilinx devices. As
expected, our level of resource utilization is between the comparison
implementations whose decoders realize serial and fully-parallel
architectures.

%% file: conclusion.tex
%\section{Conclusion} This work presented an early investigation into
%the feasibility of a hardware-based \admmlp decoder.  We target an
%\gls{fpga} platform with fixed logic, memory, and routing resources.
%We showed that an \gls{fpga}-based, partially-parallel, decoder
%architecture can be used to study \admmlp decoding performance of
%linear block codes shorter than 1000 bits.  We now mention some
%future work.  First is further investigation of error-floor regime
%performance for LP (and penalized-LP~\cite{draper_2014_penalized_lp})
%decoding.  Second, the development of simplified decoding algorithms
%that maintain error-floor performance while reducing the required
%bit-width.  Our ultimate objective is a fully-custom silicon
%integrated circuit implementation, which would be required to achieve
%decoding speedup for longer codes.  In such an implementation, it
%would be beneficial to explore new hardware architectures that would
%provide greater information throughput, and parity-check matrix
%reconfigurability.

%\section{Discussion and Conclusions}

In this paper we demonstrate that \admmlp decoding can attain
excellent error-rate performance in a fixed-point
implementation. While our initial implementation requires higher
fixed-point precision and more logic resources than the Min-Sum
algorithm, this study points to numerous possible avenues for future
developments, which could bring ADMM-LP's resource
requirements into line with those of other message-passing decoders.

One avenue is algorithmic simplification.  Just as Min-Sum can be
viewed as a computationally simple approximation of Sum-Product
\gls{bp}, we can seek approximations of \admmlp that
preserve its the high-\gls{snr} performance. As one example,
in~\cite{wasson_2016_thesis} it is observed that implementing
partial-sort (rather than full-sort) can result in a negligible
increase in error rates.

A second set of directions is hardware-centric.  Numerous interesting
challenges remain in the design of a hardware-efficient
implementation.  For example, it is not obvious how to implement
a \gls{cn} or a \gls{vn} unit that can handle multiple node degrees.
We believe that this problem can be solved through innovative hardware
sharing or algorithmic generalization.  As a second example, \admmlp
also provides an opportunity for simplifying message-passing networks,
especially when considering a fully-parallel architecture.  This is
because the same message is sent from each variable to all connected
checks.  Such message broadcasting can perhaps be exploited to reduce
interconnect complexity.  Finally, this study is a first step en-route
to the development of a fully custom, in-silicon, \gls{asic}.  An
\gls{asic} would allow for high-performance, power-optimized register
files and customized message-passing resources that would yield
significant performance improvements not possible in an \gls{fpga}
realization.

Referring to Table~\ref{table:FPGA-decoder-comparison}, we note that
while our normalized throughput per iteration is $35\times$ lower than
that of the Min-Sum decoder of~\cite{Chandrasetty2011_comp}, our
ADMM-LP decoder achieves a \gls{ber} nearly $100\times$ better.  This
is the crux of the matter.  If one is concerned with applications
where excellent performance in the high-\gls{snr} regime is required,
a regime where algorithms such as Min-Sum or Sum-Product encounter
error-floor problems, then \admmlp should be an algorithm of great
interest.  Our current implementation outperforms Min-Sum with less
than an order of magnitude difference in the number of
\gls{fpga} resources required.  Further development, and innovation,
could turn \admmlp into the algorithm of choice in such regimes of
operation.

%% file: master.bbl
\begin{thebibliography}{10}
\providecommand{\url}[1]{#1}
\csname url@samestyle\endcsname
\providecommand{\newblock}{\relax}
\providecommand{\bibinfo}[2]{#2}
\providecommand{\BIBentrySTDinterwordspacing}{\spaceskip=0pt\relax}
\providecommand{\BIBentryALTinterwordstretchfactor}{4}
\providecommand{\BIBentryALTinterwordspacing}{\spaceskip=\fontdimen2\font plus
\BIBentryALTinterwordstretchfactor\fontdimen3\font minus
  \fontdimen4\font\relax}
\providecommand{\BIBforeignlanguage}[2]{{%
\expandafter\ifx\csname l@#1\endcsname\relax
\typeout{** WARNING: IEEEtran.bst: No hyphenation pattern has been}%
\typeout{** loaded for the language `#1'. Using the pattern for}%
\typeout{** the default language instead.}%
\else
\language=\csname l@#1\endcsname
\fi
#2}}
\providecommand{\BIBdecl}{\relax}
\BIBdecl

\bibitem{berrou_1993_turbo}
C.~Berrou, A.~Glavieux, and P.~Thitimajshima, ``Near {Shannon} limit
  error-correcting coding and decoding: Turbo-codes,'' in \emph{Proc. IEEE Int.
  Conf. Comm.}, May 1993, pp. 1064--1070.

\bibitem{mackay_1995_ldpc}
D.~J.~C. MacKay and R.~M. Neal, ``Good codes based on very sparse matrices,''
  in \emph{IMA Int.~Conf.~on Cryptography and Coding}, C.~Boyd, Ed.\hskip 1em
  plus 0.5em minus 0.4em\relax Springer, Dec. 1995, pp. 100--111.

\bibitem{kschischang_2001_sum_product}
F.~R. Kschischang, B.~J. Frey, and H.-A. Loeliger, ``Factor graphs and the
  sum-product algorithm,'' \emph{{IEEE} Trans. Inf. Theory}, vol.~47, no.~2,
  pp. 498--519, Feb. 2001.

\bibitem{richardson_2003_error}
T.~Richardson, ``Error floors of {LDPC} codes,'' in \emph{Proc. Allerton Conf.
  Comm.~Control and Comp.}, vol.~41, no.~3, Oct. 2003, pp. 1426--1435.

\bibitem{feldman_2005_journal}
J.~Feldman, M.~J. Wainwright, and D.~R. Karger, ``Using linear programming to
  decode binary linear codes,'' \emph{{IEEE} Trans. Inf. Theory}, vol.~51,
  no.~3, pp. 954--972, Mar. 2005.

\bibitem{seigel_2008_adaptive_lp}
{M. H. Taghavi} and P.~H. Siegel, ``Adaptive methods for linear programming
  decoding,'' \emph{{IEEE} Trans. Inf. Theory}, vol.~54, no.~12, pp.
  5396--5410, Nov. 2008.

\bibitem{feldman_2005_constant}
J.~Feldman, T.~Malkin, R.~A. Servedio, C.~Stein, and M.~J. Wainwright,
  ``Message-passing algorithms and improved {LP} decoding,'' in \emph{Proc.\
  Int.\ Symp.\ Inf.\ Theory}, Chicago, IL, Jun. 2005.

\bibitem{arora_2009_constant}
A.~Arora, D.~Steuer, and C.~Daskalakis, ``Message-passing algorithms and
  improved {LP} decoding,'' in \emph{{ACM} Symposium on Theory of Computing
  (STOC)}, May 2009.

\bibitem{vontobel_2006_low_complex_lp}
P.~O. Vontobel and R.~Koetter, ``Towards low-complexity linear-programming
  decoding,'' in \emph{Proc. 4th Int. Symp. Turbo Codes and Related Topics},
  Munich, Germany, Apr. 2006, pp. 1--9.

\bibitem{burshtein_2009_iterative_lp}
D.~Burshtein, ``Iterative approximate linear programming decoding of {LDPC}
  codes with linear complexity,'' \emph{{IEEE} Trans. Inf. Theory}, vol.~55,
  no.~11, pp. 4835--4859, Nov. 2009.

\bibitem{draper_2013_admm_lp}
S.~Barman, X.~Liu, S.~C. Draper, and B.~Recht, ``Decomposition methods for
  large scale {LP} decoding,'' \emph{{IEEE} Trans. Inf. Theory}, vol.~59,
  no.~12, pp. 7870--7886, Dec. 2013.

\bibitem{boyd_2011_admm}
S.~Boyd, N.~Parikh, E.~Chu, B.~Peleato, and J.~Eckstein, ``Distributed
  optimization and statistical learning via the alternating direction method of
  multipliers,'' \emph{Found.~Trends Machine Learning}, no.~1, pp. 1–--122,
  2011.

\bibitem{liu_2014_instanton}
X.~Liu and S.~C. Draper, ``Instanton search algorithm for the {ADMM} penalized
  decoder,'' in \emph{Proc.\ Int.\ Symp.\ Inf.\ Theory}, Honolulu, Jun. 2014.

\bibitem{liu_2015_jumpLinear}
------, ``{ADMM} decoding on trapping sets,'' in \emph{Proc.\ Int.\ Symp.\
  Inform.\ Theory}, Hong Kong, Jun. 2015.

\bibitem{draper_2014_penalized_lp}
------, ``The {ADMM} penalized decoder for {LDPC} codes,'' \emph{{IEEE} Trans.
  Inf. Theory}, vol.~62, no.~6, pp. 2966--2984, Jun. 2016.

\bibitem{wang_2009_multistage}
Y.~Wang, J.~S. Yedidia, and S.~C. Draper, ``Multi-stage decoding of {LDPC}
  codes,'' in \emph{Proc.\ Int.\ Symp.\ Inf.\ Theory}, South Korea, Jul. 2009.

\bibitem{liu_2016_nonBinADMM}
X.~Liu and S.~C. Draper, ``{ADMM} {LP} decoding of non-binary ldpc codes in
  $\mathbb{F}_{2^m}$,'' \emph{{IEEE} Trans. Inf. Theory}, vol.~62, pp.
  2985--3010, Jun. 2016.

\bibitem{liu_2016_multiperm}
------, ``{LP}-decodable multipermutation codes,'' \emph{{IEEE} Trans. Inf.
  Theory}, vol.~62, pp. 1631--1648, Apr. 2016.

\bibitem{siegel_2013_projection_lp}
X.~Zhang and P.~H. Siegel, ``Efficient iterative {LP} decoding of {LDPC} codes
  with alternating direction method of multipliers,'' in \emph{Proc. Int. Symp.
  Inf. Theory}, Istanbul, Turkey, Jul. 2013, pp. 1501--1505.

\bibitem{kleijn_2013_pp_projection}
G.~Zhang, R.~Heusdens, and W.~B. Kleijn, ``Large scale {LP} decoding with low
  complexity,'' \emph{IEEE Comm.~Lett.}, pp. 2152--2155, Nov. 2013.

\bibitem{wasson_2015_pp}
M.~Wasson and S.~C. Draper, ``Hardware based projection onto the parity
  polytope and probability simplex,'' in \emph{Proc. 49th Asilomar Conf.
  Signals, Systems, Computers}, Pacific Grove, CA, Nov. 2015, pp. 1015--1020.

\bibitem{debbabi2016}
I.~Debbabi, B.~L. Gal, N.~Khouja, F.~Tlili, and C.~Jego, ``Fast converging
  {ADMM}-penalized algorithm for {LDPC} decoding,'' \emph{{IEEE} Commun.
  Lett.}, vol.~20, no.~4, pp. 648--651, Apr. 2016.

\bibitem{debbabi2015}
------, ``Analysis of {ADMM}-{LP} algorithm for {LDPC} decoding, a first step
  to hardware implementation,'' in \emph{IEEE Int.\ Conf.\ Electronics,
  Circuits, and Systems}, Cairo, Egypt, Dec. 2015, pp. 356--359.

\bibitem{Jiao2015}
X.~Jiao, H.~Wei, J.~Mu, and C.~Chen, ``Improved {ADMM Penalized} decoder for
  irregular low-density parity-check codes,'' \emph{{IEEE} Commun. Lett.},
  vol.~19, no.~6, pp. 913--916, Jun. 2015.

\bibitem{wei2016}
H.~Wei, X.~Jiao, and J.~Mu, ``Reduced-complexity linear programming decoding
  based on {ADMM} for {LDPC} codes,'' \emph{{IEEE} Commun. Lett.}, vol.~19,
  no.~6, pp. 909--912, Jun. 2015.

\bibitem{tanner_2001_refcode}
R.~M. Tanner, D.~Sridhara, and T.~Fuja, ``A class of group-structured {LDPC}
  codes,'' in \emph{Proc.\ {ICSTA}}, Ambleside, U.K., Jul. 2001.

\bibitem{IEEE-80211ad_rev}
\emph{IEEE Std 802.11ad-2012 (Amend.~to IEEE Std 802.11-2012, as amended by
  IEEE Std 802.11ae-2012 and 802.11aa-2012)}, pp. 1--628, Dec 2012.

\bibitem{feldman_2003_phd}
J.~Feldman, ``Decoding error-correcting codes via linear programming,'' Ph.D.
  dissertation, Massachusetts Institute of Technology, USA, 2003.

\bibitem{wasson_2016_thesis}
M.~Wasson, ``Hardware-based linear program decoding with the alternating
  direction method of multipliers,'' Master's thesis, University of Toronto,
  Canada, Nov. 2016.

\bibitem{barman_2011}
S.~Barman, X.~Liu, S.~C. Draper, and B.~Recht, ``Decomposition methods for
  large scale {LP} decoding,'' in \emph{Proc. Allerton Conf. Comm. Control
  Computing}, Monticello, IL, Sep. 2011.

\bibitem{zhang_2012_phd}
X.~Zhang, ``{LDPC} codes: {S}tructural analysis and decoding techniques,''
  Ph.D. dissertation, University of California, San Diego, USA, 2012.

\bibitem{siegel_2012_cut_search}
X.~Zhang and P.~H. Siegel, ``Adaptive cut generation algorithm for improved
  linear programming decoding of binary linear codes,'' \emph{{IEEE} Trans.
  Inf. Theory}, vol.~58, no.~10, pp. 6581--6594, Oct. 2012.

\bibitem{duchi_2008_simplex_projection}
J.~Duchi, S.~Shalev-Shwartz, Y.~Singer, and T.~Chandra, ``Efficient projections
  onto the $\ell_1$-ball for learning in high dimensions,'' in \emph{Proc. Int.
  Conf. Machine Learning}, San Diego, USA, Dec. 2008.

\bibitem{ss_2006_simplex}
S.~Shalev-Shwartz and Y.~Singer, ``Efficient learning of label ranking by soft
  projections onto polyhedra,'' \emph{Journal of Machine Learning Research},
  vol.~7, pp. 1567–--1599, Jul. 2006.

\bibitem{kou_2001_qc}
Y.~Kou, S.~Lin, and M.~P.~C. Fossorier, ``Low-density parity-check codes based
  on finite geometries: A rediscovery and new results,'' \emph{{IEEE} Trans.
  Inf. Theory}, vol.~47, no.~7, pp. 2711--2736, Nov. 2001.

\bibitem{fossorier_2004_qc}
M.~P.~C. Fossorier, ``Quasicyclic low-density parity-check codes from circulant
  permutation matrices,'' \emph{{IEEE} Trans. Inf. Theory}, vol.~50, no.~8, pp.
  1788--1793, Aug. 2004.

\bibitem{Hocevar2004}
D.~E. Hocevar, ``A reduced complexity decoder architecture via layered decoding
  of {LDPC} codes,'' in \emph{Proc.\ Work. Signal Proc.~Sys.}, Oct. 2004.

\bibitem{Park2014}
Y.~S. Park, D.~Blaauw, D.~Sylvester, and Z.~Zhang, ``Low-power high-throughput
  {LDPC} decoder using non-refresh embedded {DRAM},'' \emph{IEEE J. Solid-State
  Circuits}, vol.~49, no.~3, pp. 783--794, Mar. 2014.

\bibitem{knuth_1998_tacp}
D.~E. Knuth, \emph{The Art of Computer Programming: Sorting and Searching},
  2nd~ed.\hskip 1em plus 0.5em minus 0.4em\relax Redwood City, USA: Addison
  Wesley Longman, 1998, vol.~2.

\bibitem{ladner_1980_prefix}
R.~E. Ladner and M.~J. Fischer, ``Parallel prefix computation,'' \emph{J. of
  the ACM}, vol.~27, no.~4, pp. 831--838, Oct. 1980.

\bibitem{liu_2015_rng}
\BIBentryALTinterwordspacing
G.~Liu. (2015) Gaussian noise generator. OpenCores. [Online]. Available:
  \url{http://opencores.org/project,gng}
\BIBentrySTDinterwordspacing

\bibitem{draper_2013_qc}
Y.~Wang, S.~C. Draper, and J.~S. Yedidia, ``Hierarchical and high-girth {QC}
  {LDPC} codes,'' \emph{{IEEE} Trans. Inf. Theory}, vol.~59, no.~7, pp.
  4553--4583, Jul. 2013.

\bibitem{khan2011digital}
S.~A. Khan, \emph{Digital design of signal processing systems: a practical
  approach}.\hskip 1em plus 0.5em minus 0.4em\relax John Wiley \& Sons, 2011.

\bibitem{liu_2013_dp}
\BIBentryALTinterwordspacing
X.~Liu. (2015) {ADMM} decoder. [Online]. Available:
  \url{https://sites.google.com/site/xishuoliu/codes}
\BIBentrySTDinterwordspacing

\bibitem{siegel_2014_bp}
B.~K. Butler and P.~H. Siegel, ``Error floor approximation for {LDPC} codes in
  the {AWGN} channel,'' \emph{{IEEE} Trans. Inf. Theory}, vol.~60, no.~12, pp.
  7416--7441, Dec. 2014.

\bibitem{Lei2006_comp}
X.~Lei, T.~Zhenhui, and Y.~Dongping, ``The moderate-throughput and
  memory-efficient {LDPC} decoder,'' in \emph{2006 8th International Conference
  on Signal Processing}, vol.~3, 2006.

\bibitem{Zarubica2007_comp}
R.~Zarubica, S.~G. Wilson, and E.~Hall, ``Multi-{Gbps} {FPGA}-based low density
  parity check ({LDPC}) decoder design,'' in \emph{in Proc. Global Telecomm.
  Conf.}, Nov 2007, pp. 548--552.

\bibitem{Bhagawat2005_comp}
P.~Bhagawat, M.~Uppal, and G.~Choi, ``{FPGA} based implementation of decoder
  for array low-density parity-check codes,'' in \emph{in Proc. Int. Conf.
  Acoustics, Speech, and Signal Proc.}, Mar. 2005.

\bibitem{Chandrasetty2011_comp}
V.~A. Chandrasetty and S.~M. Aziz, ``A multi-level hierarchical quasi-cyclic
  matrix for implementation of flexible partially-parallel {LDPC} decoders,''
  in \emph{Int.\ Conf.\ Multimedia, Expo}, July 2011, pp. 1--7.

\bibitem{alteraALMxilinxSlice}
\BIBentryALTinterwordspacing
``Stratix {II} vs. {V}irtex-4 density comparison,'' Altera White Paper, Aug.
  2002. [Online]. Available:
  \url{https://www.altera.com/en\_US/pdfs/literature/wp/wpstxiixlnx.pdf}
\BIBentrySTDinterwordspacing

\end{thebibliography}
